\documentclass[twocolumn]{aastex62}
\pdfoutput=1 
\usepackage{amsmath,amstext}
\usepackage{CJK}
\usepackage{float}
\usepackage{multirow}
\usepackage[T1]{fontenc}
\usepackage[figure,figure*]{hypcap}
\usepackage{threeparttable}
\usepackage{amsmath}	
\usepackage{amssymb}	
\usepackage{cases}
\usepackage{longtable}
\usepackage{xcolor}
\usepackage{booktabs}
\usepackage{cleveref}
\usepackage{paralist}
\graphicspath{{fig/}}

\newcommand {\Msun}{{M$_{\odot}$}}

\PassOptionsToPackage{pdfpagelabels=false}{hyperref}
\PassOptionsToPackage{nonamebreak}{natbib}   
\shorttitle{Precise Ages of Field Stars}
\shortauthors{DQ, et al.}

\begin{document}
\begin{CJK}{UTF8}{gbsn}
\setlength{\LTleft}{0pt} \setlength{\LTright}{0pt} 
\title{Precise Ages of Field Stars From White Dwarf Companions in Gaia DR2}
\correspondingauthor{Ha-ijun Tian}
\email{hjtian@lamost.org}
\author{Dan Qiu (邱丹)}
\affil{China Three Gorges University, Yichang 443002, China}
\affil{Center of Astronomy and Space Science Research, China Three Gorges University, Yichang 443002, China} 
\author{Hai-Jun Tian (田海俊)}
\affil{China Three Gorges University, Yichang 443002, China}
\affil{Center of Astronomy and Space Science Research, China Three Gorges University, Yichang 443002, China} 
\author{Xi-Dong Wang (王习东)}
\affil{China Three Gorges University, Yichang 443002, China}
\author{Jia-Lu Nie (聂嘉潞)}
\affil{Key Lab for Optical Astronomy, National Astronomical Observatories, Chinese Academy of Sciences, Beijing 100012, China}
\author{Ted von Hippel}
\affil{Physical Sciences Department, Embry-Riddle Aeronautical University, Daytona Beach, FL 32114, USA}
\author{Gao-Chao Liu (刘高潮)}
\affil{China Three Gorges University, Yichang 443002, China}
\affil{Center of Astronomy and Space Science Research, China Three Gorges University, Yichang 443002, China} 
\author{Morgan Fouesneau}
\affil{Max Planck Institute for Astronomy, K\"onigstuhl 17, D-69117 Heidelberg, Germany}
\author{Hans-Walter Rix}
\affil{Max Planck Institute for Astronomy, K\"onigstuhl 17, D-69117 Heidelberg, Germany}

\begin{abstract}

We analyze 4\,050 wide binary star systems involving a white dwarf (WD) and usually a main sequence (MS) star, drawn from the large sample assembled by  \citet[][hereafter, T20]{Tian_2020}. Using the modeling code BASE-9, we determine the system's ages, the WD progenitors' ZAMS masses, the extinction values ($A_V$), and the distance moduli. Discarding the cases with poor age convergences, we obtain ages for 3\,551 WDs, with a median age precision of $\sigma_{\tau}/\tau = 20$\%, and system ages typically in the range of 1-6 Gyr. We validated these ages against the very few known clusters and through cross-validation of 236 WD-WD binaries.
Under the assumption that the components are co-eval in a binary system, this provides precise age constraints on the usually low-mass MS companions, mostly inaccessible by any other means. 

 \end{abstract}
 
 

\section{Introduction}
\label{sect:intro}

 Stellar age is one of most fundamental ingredients to understand stellar  evolution and the assembly history of our Galaxy \citep[e.g.][]{Soderblom_2010, BW2016, Liu2017, Tian_2018}. It is therefore of great importance to determine ages of individual stars that are both precise and accurate, particularly for long-lived low-mass stars that sample the entire Milky Way formation history. Yet, stellar ages are hard to determine \citep{Soderblom_2010}, since they cannot be measured directly like other fundamental quantities, only estimated through mostly model-dependent or other empirical methods. 

There are different approaches to determine stellar ages, each with a distinct range of applicability; no single approach works well for a broad range of stellar types. The perhaps most common technique is based on the comparison of a star's position in the Hertzsprung-Russell diagram (HRD) with the theoretical predictions of stellar evolutionary models \citep[e.g.][]{Haywood_2013, Bergemann_2014}. This method is precise only for those stars near the main sequence turn-off (MSTO) or the subgiant branch where isochrones of different ages are widely separated \citep{xiang2017}. By contrast, the technique performs poorly for cool, low-mass main-sequence or giant stars, because isochrones are very close for these stars at different ages. Asteroseismology \citep{Cunha_2007} can obtain stellar ages with uncertainties at the level of about 10-20\% \citep[e.g.][]{Christensen_2009,Mazumdar_2005}, but the method is currently only applicable to stars with a limited range of spectral types that exhibit prominent solar-like oscillations \citep[see e.g.][]{Vauclair_2009}. Gyrochronology empirically relies on rotation periods and colors to derive an age, which limits this method to favor solar-type and young stars, particularly stars with spots which make it easy to measure the rotation periods \citep[e.g.][]{Barnes_2007,Angus_2015}. Some other empirical methods rely on the stellar activity, e.g., helium flash \citep{Tian_2017_helium} and depletion of lithium\citep[e.g.][]{Skumanich_1972,Mentuch_2008,Nascimento_2009}, or element abundances, e.g., carbon and nitrogen abundances \citep{Martig_2016}. These approaches are largely restricted to stars of $\geq$ 1 \Msun, the typical uncertainty of age derived from these methods is about 30\% \citep{Chaplin_2014}, and in some cases even larger than 40\% \citep{Martig_2016}.

It has long been appreciated that white dwarfs (WDs) can serve as precise stellar clocks, since they have well-understood cooling models \citep{Bergeron_1995}. Total age estimates need to combine the cooling age with the precursor age, which requires constraints on the initial-final mass relations (IFMR, \citet{Weidemann_1977,Weidemann_2000,catal_2008a}). With a self-consistent and robust Bayesian statistical approach \citep[BASE-9][]{Hipple_2006}, a precise age (with a typical uncertaity of 10-15\%) can be derived for individual WDs by requiring only quality multi-band photometry and precise distances \citep{Malley_2013, Webster_2015}.
If WDs are in (resolved) binaries with stars, in particular MS stars, they may serve to age-date their companions, as it is a safe assumption that the binary components
are co-eval. In this way, \citet{Catal_2008}, \citet{Garc_2011}, \citet{Zhao_2011}, and \citet{ Rebassa-Mansergas_2016} estimated ages for 6, 27, 36, and 23 MS-WD binaries, respectively. The excellent parallax and proper motions provided by Gaia allow us to significantly extend such binary samples. As a pilot study, \citet{Fouesneau2019} identified nearly 100 candidates for wide MS-WD binary systems: each contained a faint WD whose GPS1 proper motion \citep{Tian_2017} matched that of a brighter MS star in Gaia/TGAS \citep{gaia2016}, then they estimated ages for these binaries with a Bayesian approach \citep[see also][]{Malley_2013}. 

In this work, we build the current largest MS-WD binary catalog with Gaia DR2 \citep{gaia2018}, and determine the ages of these systems. Gaia DR2 provides us more than 1.3 billion stars brighter than 20.7\,mag in the G-band with measured positions, proper motions, parallaxes and colors with unprecedented precision. We selected stars whose positions, proper motions, and parallaxes were consistent with being gravitationally bound, drawing on the wide binary catalog of T20. Based on this catalog, we can easily obtain  MS-WD binaries, and then derive the precise ages for these binaries following \citet{Malley_2013} and \citet{Fouesneau2019}.
 
This paper is organized as follows: Section 2 presents the selection of MS-WD and WD-WD binary systems. In Section 3, we describe the determination of evolutionary models, prior values and photometric bands for WDs. In Section 4, we analyze the posterior results of WDs. We summarize the method and results in Section 5.


\section{Data Selection}
\label{sect:data}

We start with the catalog of initial wide binary candidate released by T20, which contains 807,611 candidates, selected from Gaia DR2 within a distance of 4.0\,kpc and a maximum projected separation $s = 1.0$\,pc. The selection criteria are summarized in Section 2 of T20. The clusters, groups, and pairs in high density regions have been removed from these candidates, mainly leaving MS-WD, WD-WD, MS-MS binaries and contaminants (a few sub-giants and giants are included in T20). Note that the contamination rate changes dramatically with projected separation, from negligibly small to dominant, particularly at $s>30\,000$\,AU as shown in Figure 4 of T20.

Our subsequent analysis also requires constraints on the interstellar extinction. 
We start with the reddening value $\rm E(B-V)$ for each star from the three-dimensional (3D) dust map of \citet{Green_2019}, and then obtain the extinction value ($A_V$) with $R_V=3.1$. Unfortunately, the 3D map of \citet{Green_2019} only covers the sources in three-quarters of the sky (i.e., $\rm decl.>-30$\degr). We therefore obtain $A_V$ for the sources with $\rm decl.<-30$\degr\ from the map of \citet[][hereafter SFD]{Schlegel_1998}. However, the SFD map provides total extinction values which are column integrated in the line-of-sight directions through the Galaxy \citep{Tian_2014}. This will overestimate the extinction for our nearby sample, particularly for the sources in the disk plane (e.g., $|b|<10\degr$). Hence, for the sources with $\rm decl.<-30$\degr\ and $|b|<10\degr$, we estimate their $A_V$ with an empirical model, i.e., the V band extinction $A_V$ is approximately 0.85 $\times\, d$, where $d$ is the distance of a source from us in kpc, since the interstellar extinction in the V band is around 0.7-1.0 $mag/kpc$ in the solar neighborhood \citep{Milne_1980, Wang_2017}. An average value (i.e, 0.85 $mag/kpc$) is adopted in this study \citep{Tian2020,Liu_2020}. Further, we transfer each $A_V$ into the extinctions in {\it Gaia} bands. \citet{Danielski_2018} introduced an approach to deduce the coefficients ($A_M$/$A_V$) and implemented in \citet{Collaboration_2018}, where $M$ can be $G$, $G_{BP}$, $G_{RP}$. The extinction coefficients of {\it Gaia} bands were defined as functions of the extinction in V-band ($A_V$) and the stellar effective temperature (in the term of the color index ($G_{BP}-G_{RP}$)). This approach effectively eliminates the deviation in the shape of the reddened isochrone if a fixed relation $A_M$/$A_V$ is adopted.
 
\subsection{Outlier Cleaning}
\label{Outlier Cleaning}

In this work, we are only interested in the MS-WD and WD-WD binaries, so we eliminate MS-MS binary candidates, i.e., we only keep the candidates with binary\_type = 1 (MS-WD) and binary\_type = 2 (WD-WD), which results in 42,839 MS-WD and 440 WD-WD candidates. 

\begin{figure}[!t]
\centering
\includegraphics[width=0.41\textwidth, trim=0.0cm 3.3cm 0.0cm 0.0cm, clip]{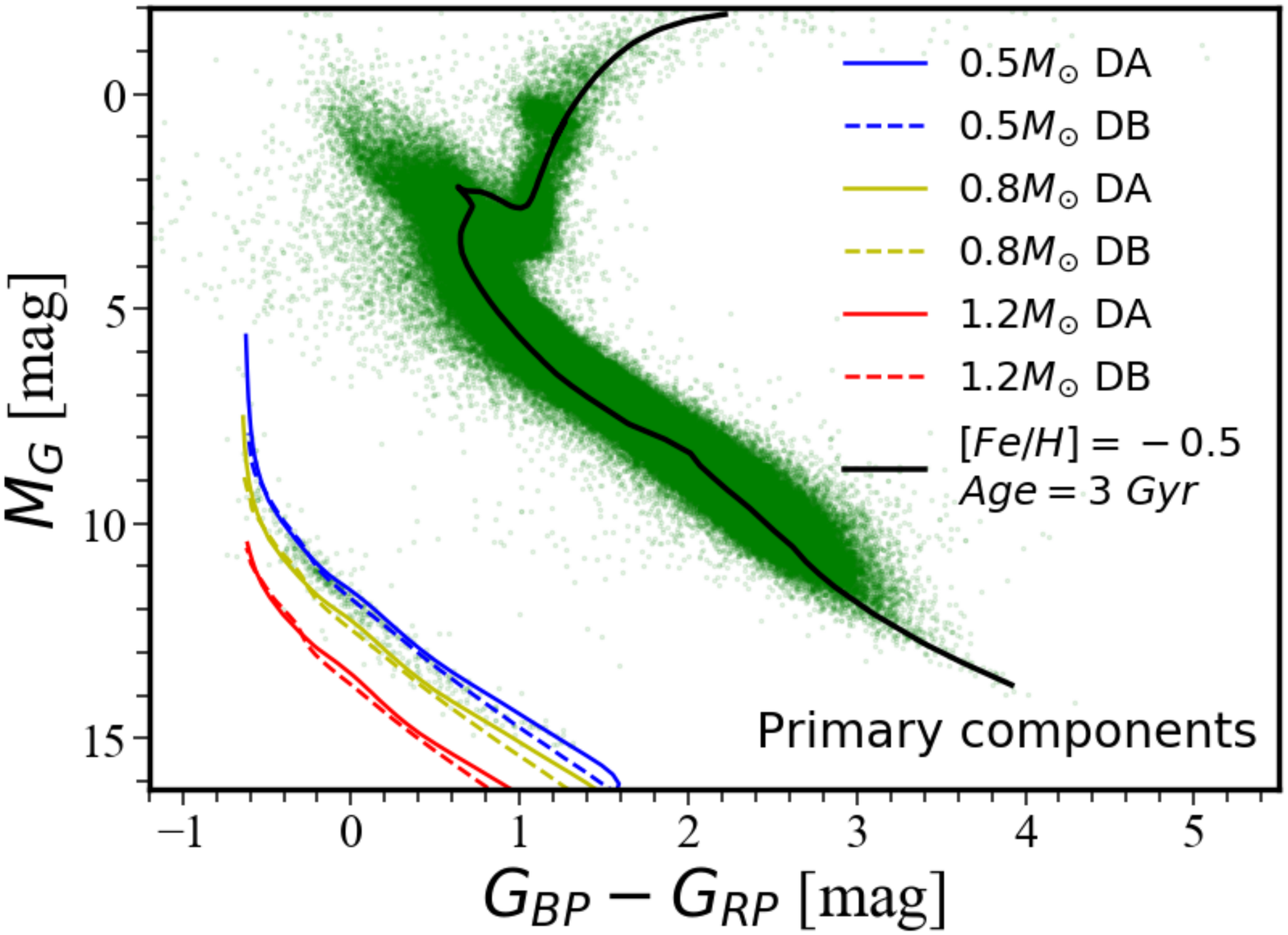}
\includegraphics[width=0.41\textwidth, trim=0.0cm 0.0cm 0.0cm 0.0cm, clip]{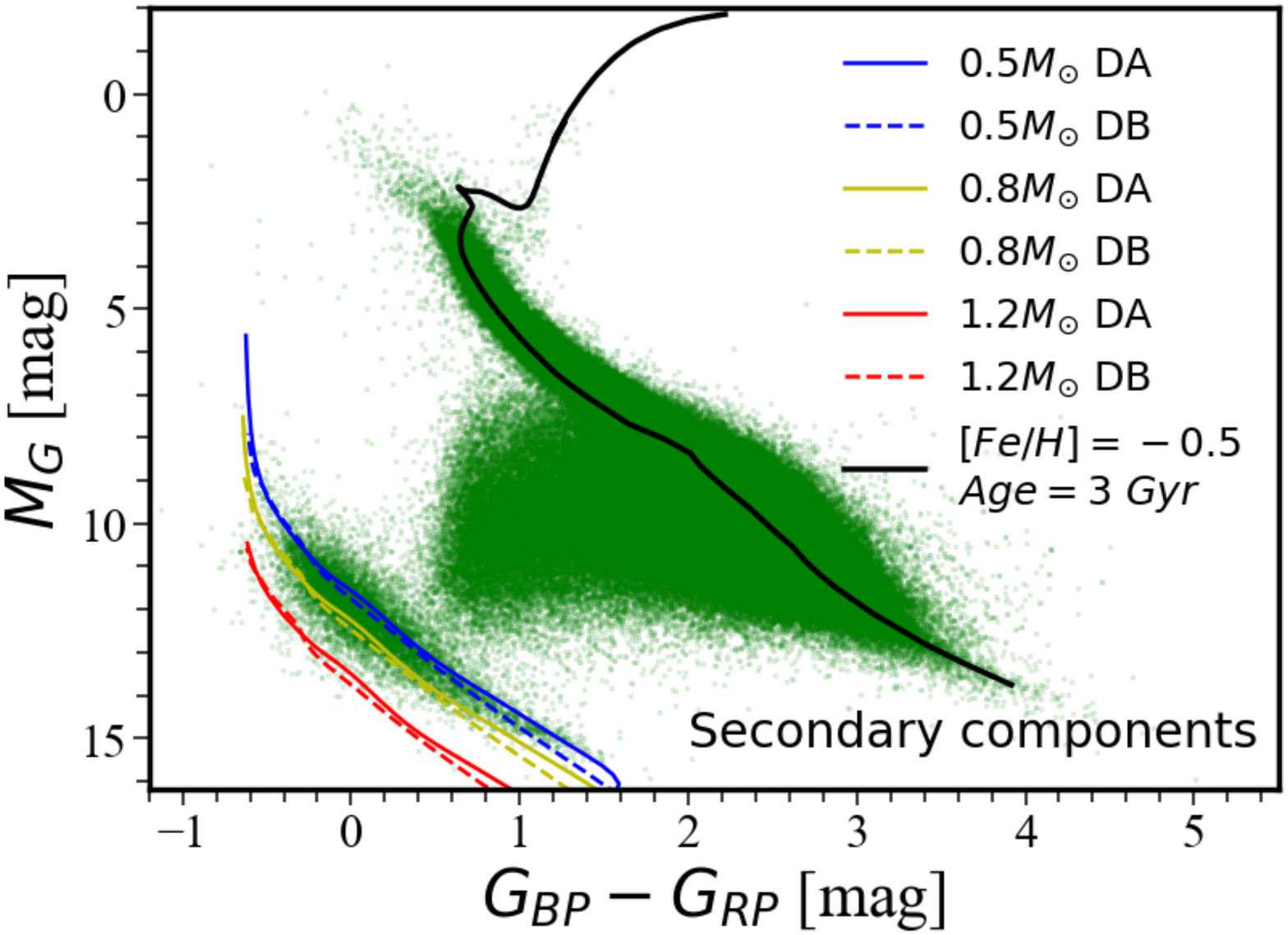}
\caption{Color-(absolute) magnitude diagrams (CMDs) for wide binary candidates. The stellar extinction is calibrated for each star. The green dots are 625,620 pairs of binary candidates, which are obtained from the initial binary catalog of T20 and covered by the 3D dust map of \citet{Green_2019}. The cooling curves of WDs with 0.5\,\Msun\ (blue), 0.8\,\Msun\ (yellow) and 1.2\Msun\ (red), and an isochrone of MS with [Fe/H]=0 and age=3\,Gyr (black) are presented for both the primary (top) and secondary (bottom) stars. The DA and DB WDs are illustrated with the solid and dashed curves. Here the ``primary components" and ``secondary components" label the brighter and fainter components, respectively.
}\label{fig:Av<3_Color}
\end{figure}

Objects with a low parallax signal-to-noise ratio (e.g., {parallax\_over\_error} $>2$) often occupy the region between the MS and WD as shown in Figure \ref{fig:Av<3_Color} (bottom). In order to obtain the binary candidates with plausible HRD positions of the WD component, we only retain candidate WDs (cWDs) of which the masses are located between the cooling curves that correspond to 0.5\,$M_{\odot}$ (the blue curves) and 1.2\,$M_{\odot}$ (the red curves), as shown in Figure \ref{fig:Av<3_Color}. For a WD with mass less than 0.5\,$M_{\odot}$, the predicted age would be inconsistent with single-star evolution during the age of the universe; WDs of such low masses are presumably results of binary star evolution. On the opposite end of the range, 1.2\,$M_{\odot}$ is close to the mass upper limit (i.e., 1.44\,$M_{\odot}$) of a WD. Because the extinction values are over-estimated for most of the nearby sample by SFD, and under-estimated for the distant sources (e.g., some giant stars) by the empirical model, we only display 625,620 binary candidates (green dots) having the 3D extinction values of \citet{Green_2019} in Figure \ref{fig:Av<3_Color}.

In addition, we remove the WD candidates outside the region of $\rm M_G> 5$ and $\rm (G_{BP}-G_{RP})< 1.7$ according to \citet{Fusillo_2019}, here ${\rm M_{G}=G+5\log\left(\varpi/mas\right)-10}$. Finally, we obtain 9589 MS-WD (215 cWDs are from the primaries, and the remaining 9,374 cWDs are from the secondaries) and 307 WD-WD candidate binaries. 

It is worth mentioning that the outliers are here removed only for the cWD components to ensure cWDs with high purity and reasonable masses, but we do not remove the contaminators for the candidate binaries. This means that the contamination rate of the candidate binaries is still high at large separations, e.g., $s>30000$\,AU, as described in Section \ref{sect:properties}; we will address that in the subsequent analysis. In the next section, we focus on the age estimation just for the WD components. The age of a MS is then equated to its WD companion if the candidate binary is a real physical binary.

\begin{figure}[!t]
\centering
\includegraphics[width=0.25\textwidth, trim=0.4cm 0.0cm 0.0cm 0.0cm, clip]{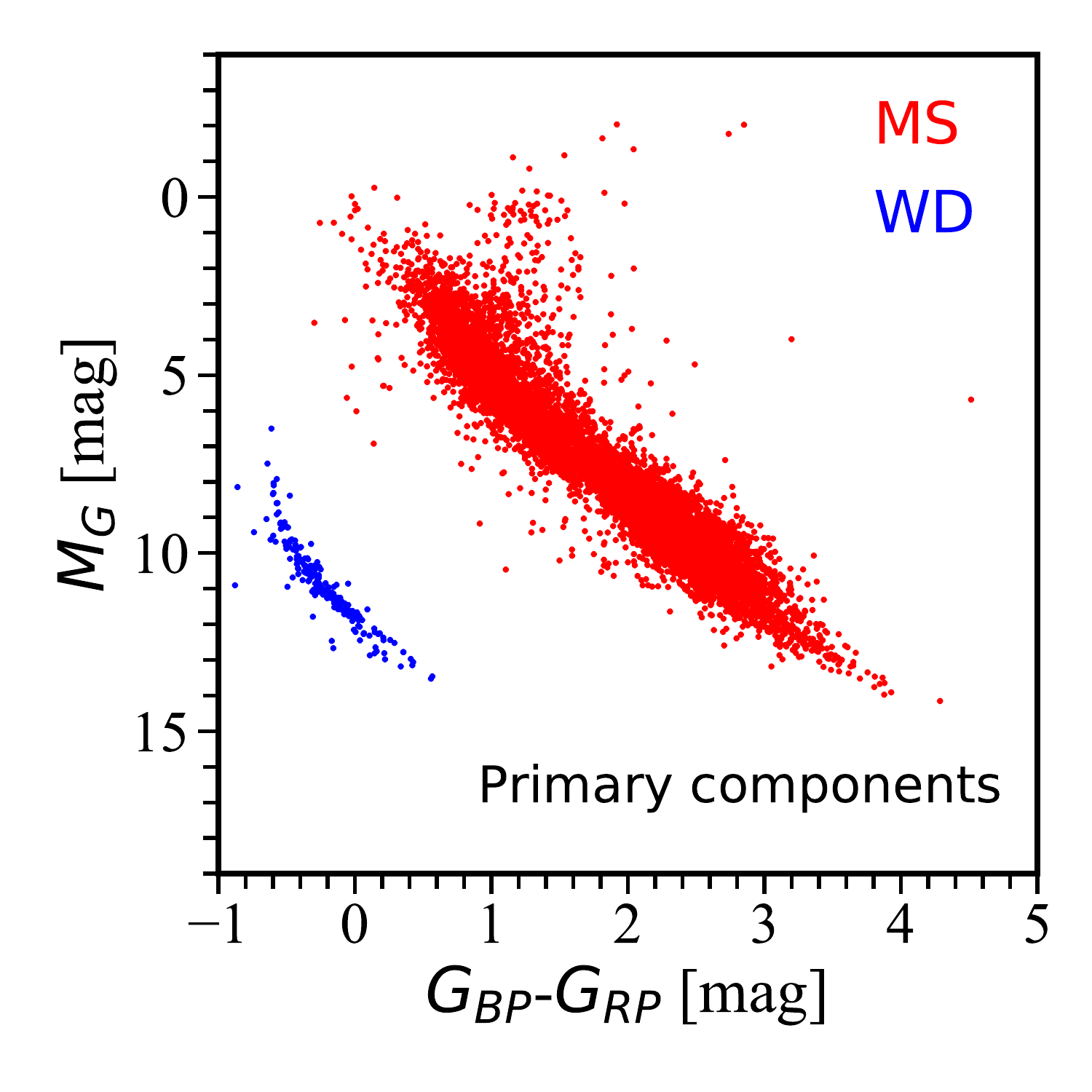}
\includegraphics[width=0.213\textwidth, trim=2.62cm 0.0cm 0.0cm 0.0cm, clip]{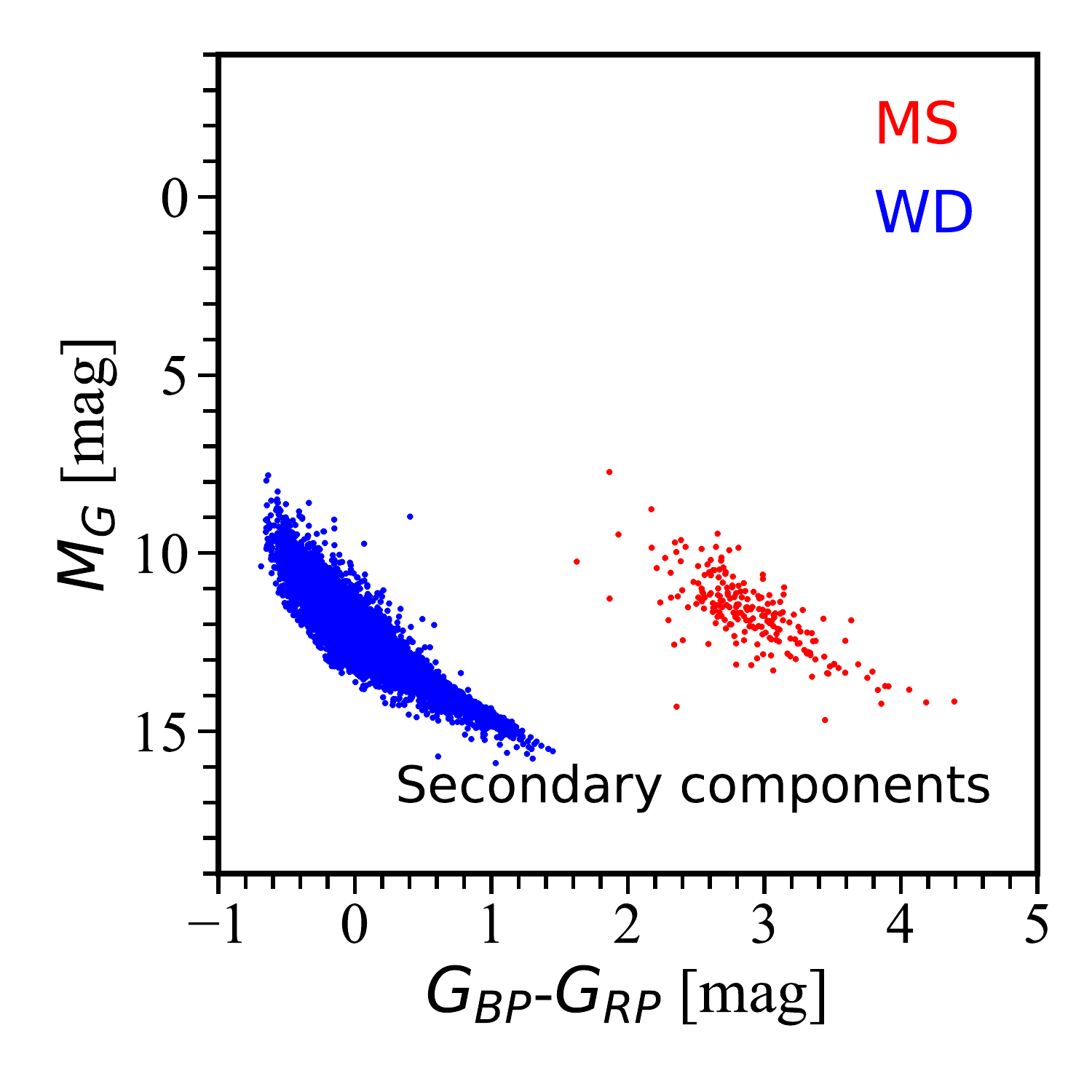}
\includegraphics[width=0.25\textwidth, trim=0.4cm 0.0cm 0.0cm 0.0cm, clip]{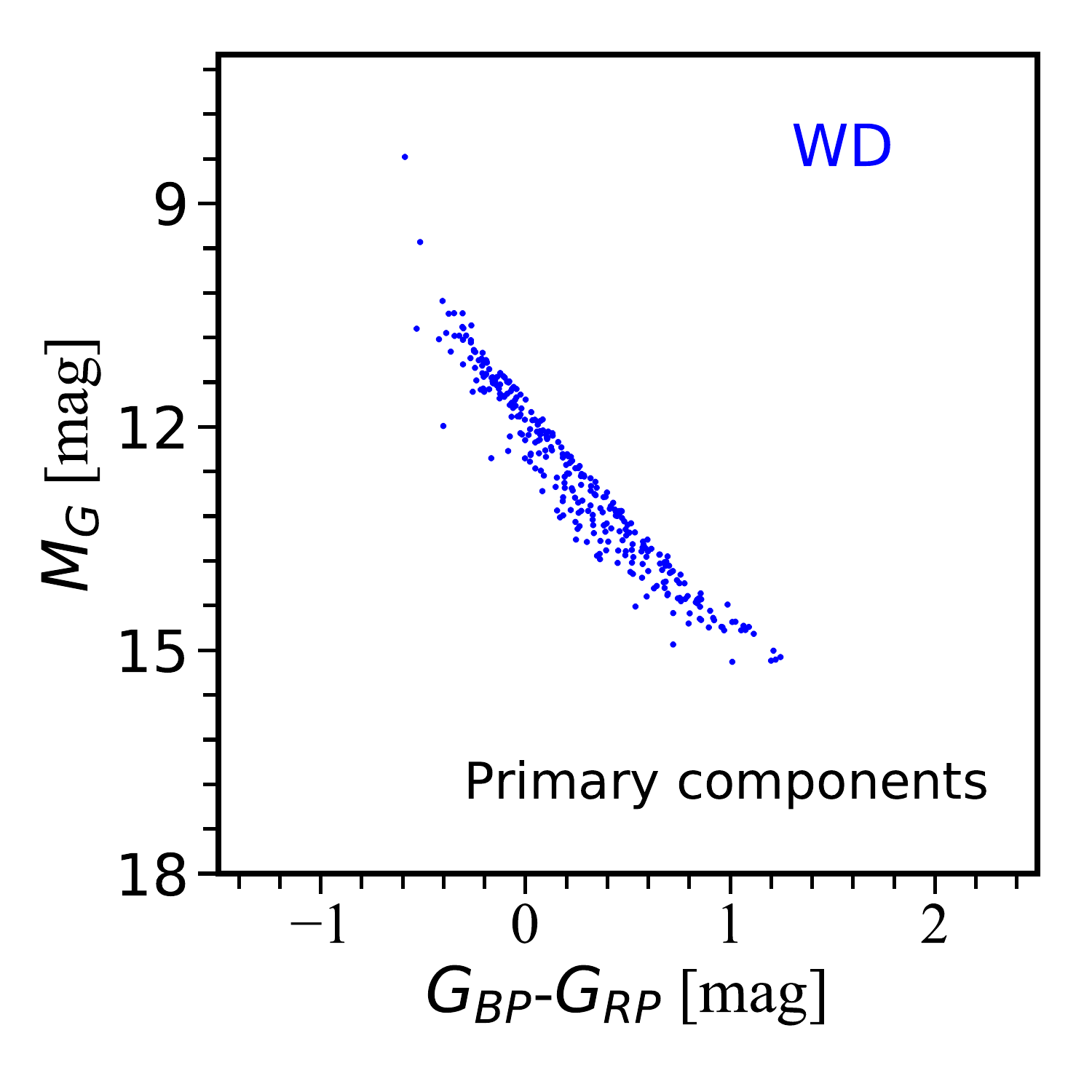}
\includegraphics[width=0.213\textwidth, trim=2.62cm 0.0cm 0.0cm 0.0cm, clip]{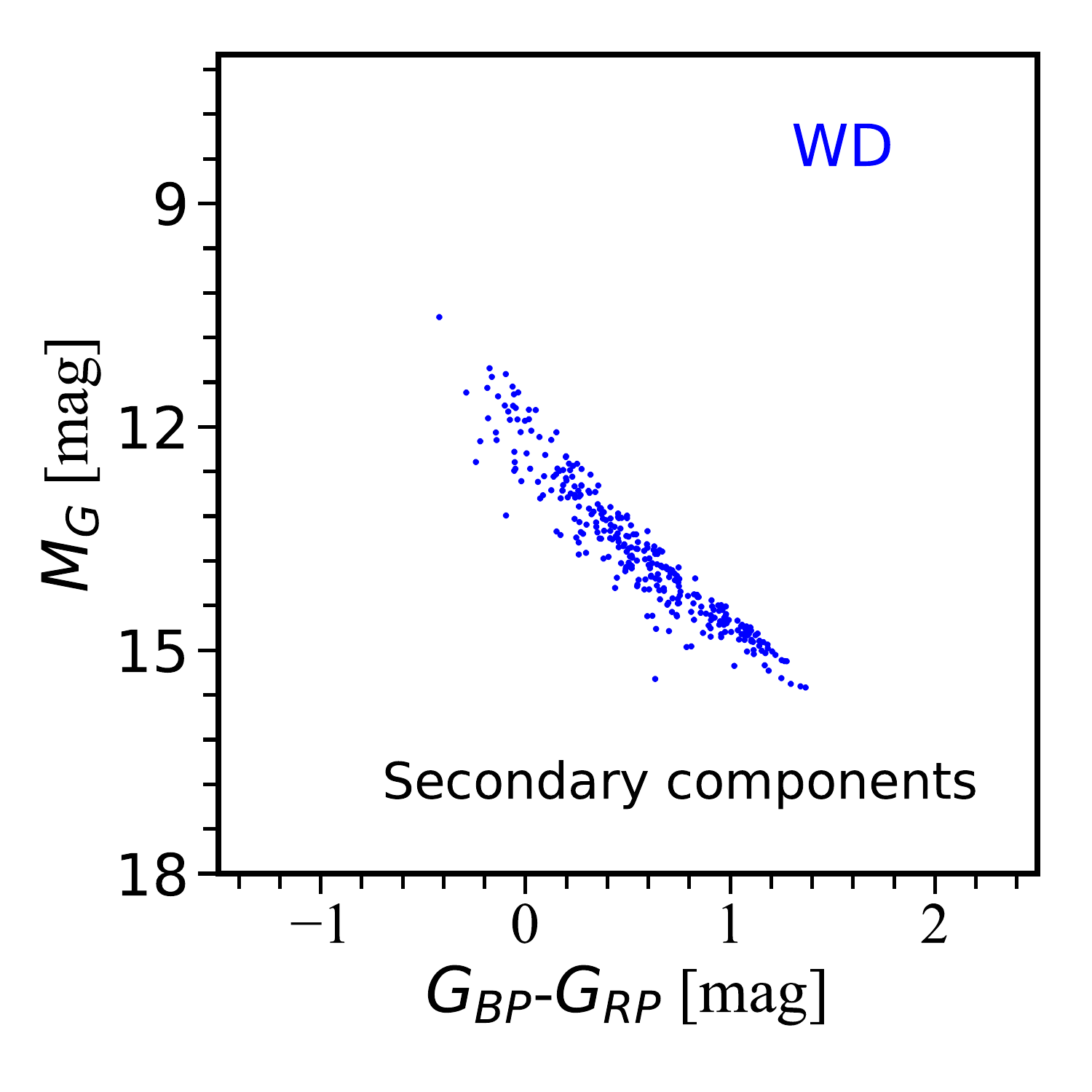}
\caption{CMDs for the 9589 MS-WD (top) and 307 WD-WD (bottom) wide binary candidates selected in Section \ref{sect:data}. In the MS-WD candidates, 215 WDs are from the primary components (top-left), and the remaining 9374 WDs are the secondary components (top-right). The MS and WD are color-coded with red and blue points, respectively.
}\label{fig:Color}
\end{figure}

\subsection{Properties of the MS-WD and WD-WD candidates}
\label{sect:properties}

We now have two samples: one includes 9,589 MS-WD binary candidates, which could be used for deriving the ages of field MS stars from their WD companions, the other consists of 307 WD-WD binary candidates, which will be used for cross-validating the ages derived in Section \ref{Age validation}. Before using the two samples, we characterize their properties, and quantify the contamination rates of the candidate binaries.

Figure \ref{fig:Color} illustrates the color-(absolute) magnitude diagrams (CMDs) for the MS-WD (top) and WD-WD (bottom) candidate binaries. The MS and WD candidates are color-coded with the red and blue dots, respectively. In the MS-WD data-set, only 215 WDs belong to the primary components (the blue dots in the top-left sub-panel), while most of WDs are the secondary components (9374 blue dots in the top-right sub-panel). This is reasonable because WDs are usually fainter than MS stars.

\begin{figure}[!t]
\centering
\includegraphics[width=0.45\textwidth, trim=0.0cm 4.5cm 0.0cm 0.0cm, clip]{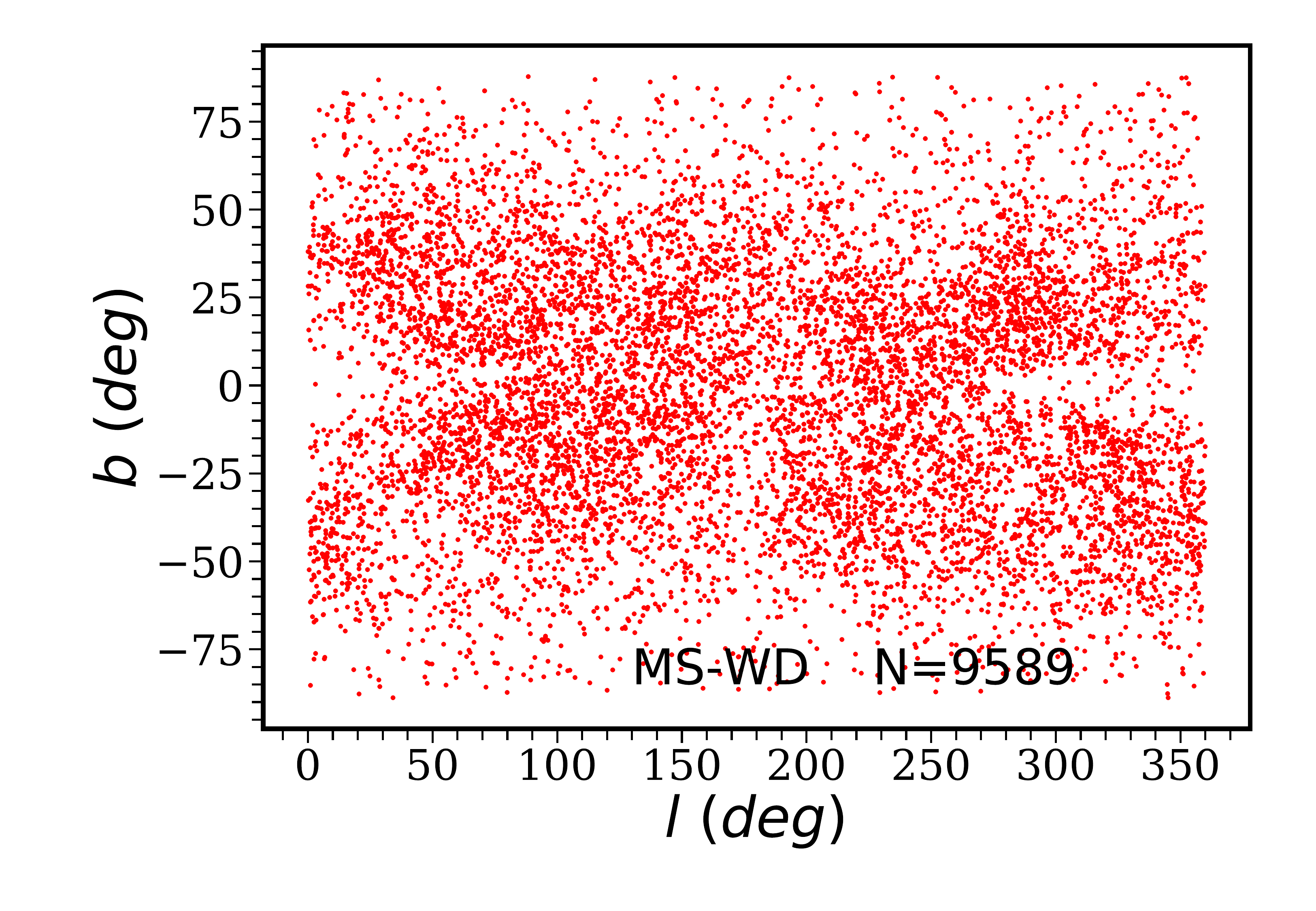}
\includegraphics[width=0.45\textwidth, trim=0.0cm 0.0cm 0.0cm 0.0cm, clip]{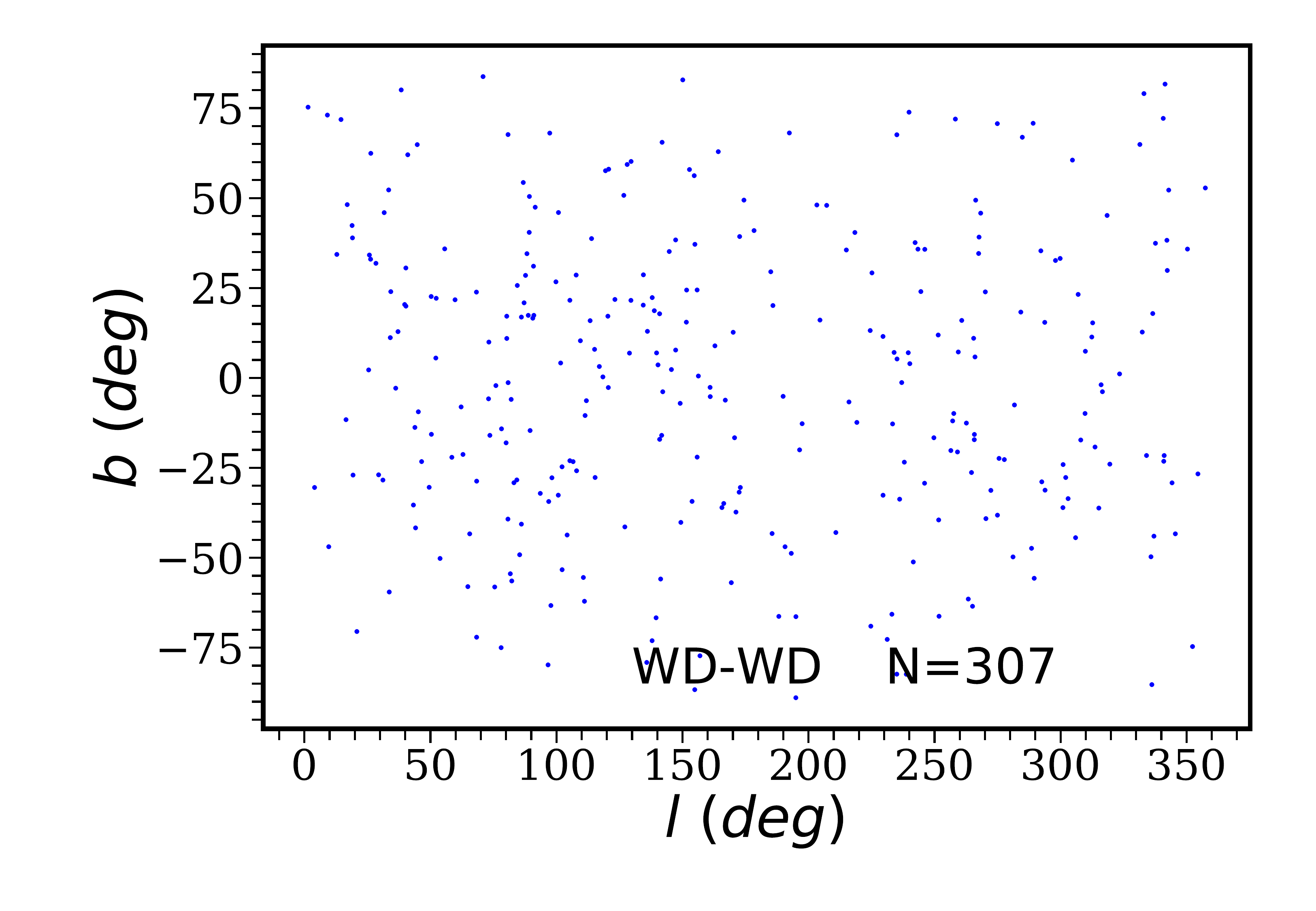}
\caption{The sky distributions ($b$ vs. $l$) for the 9589 MS-WD (top) and 307 WD-WD (bottom) candidate binaries. 
}\label{fig:bvs.l}
\end{figure}

Figure \ref{fig:bvs.l} displays the sky distributions ($b$ vs. $l$) of the MS-WD (top) and WD-WD (bottom) candidates, where $b$ and $l$ represent Galactic latitude and longitude. 

\begin{figure*}
\centering
\includegraphics[width=0.241\textwidth, trim=1.0cm 0.0cm 0.0cm 0.0cm, clip]{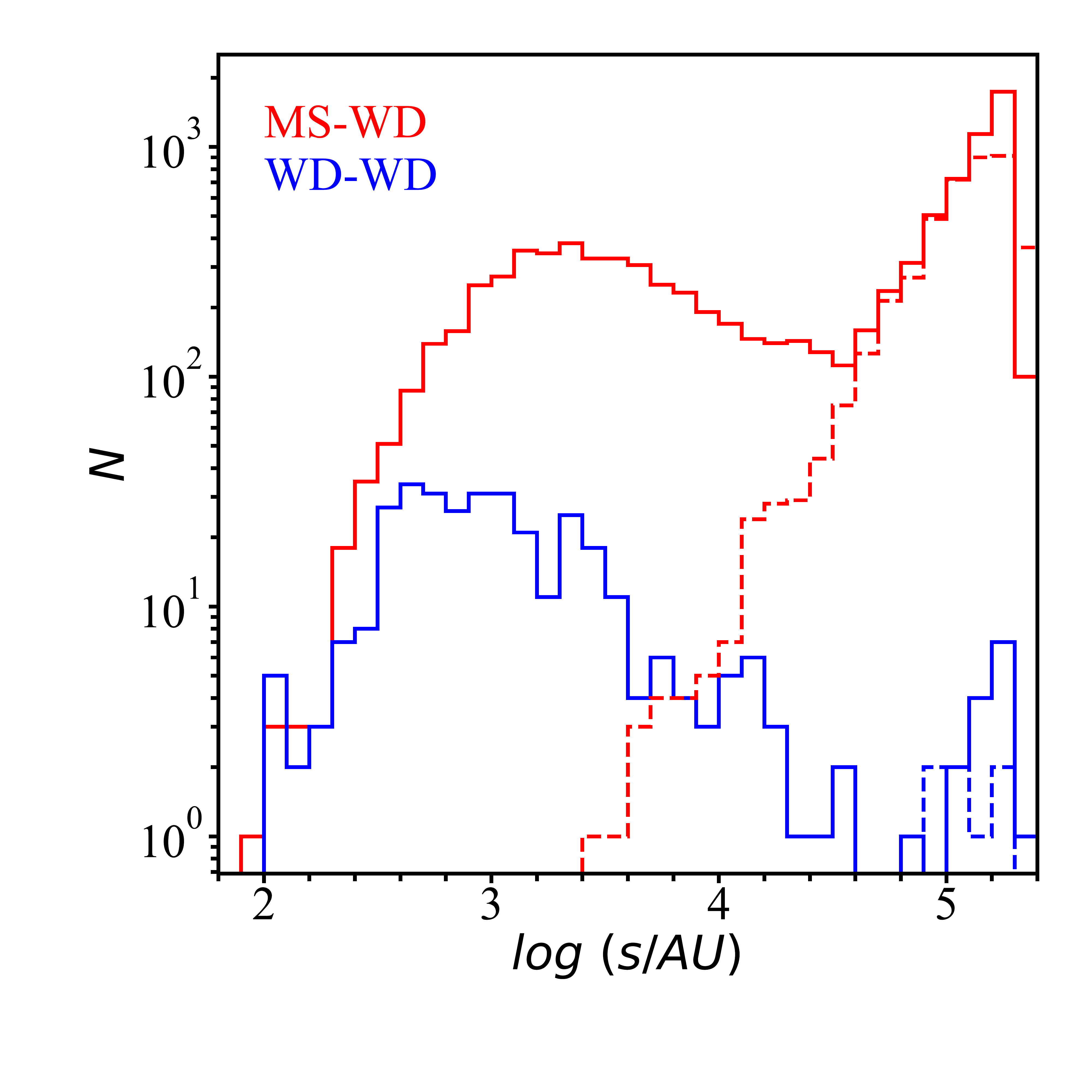}
\includegraphics[width=0.22\textwidth, trim=4.6cm 0.0cm 0.0cm 0.0cm, clip]{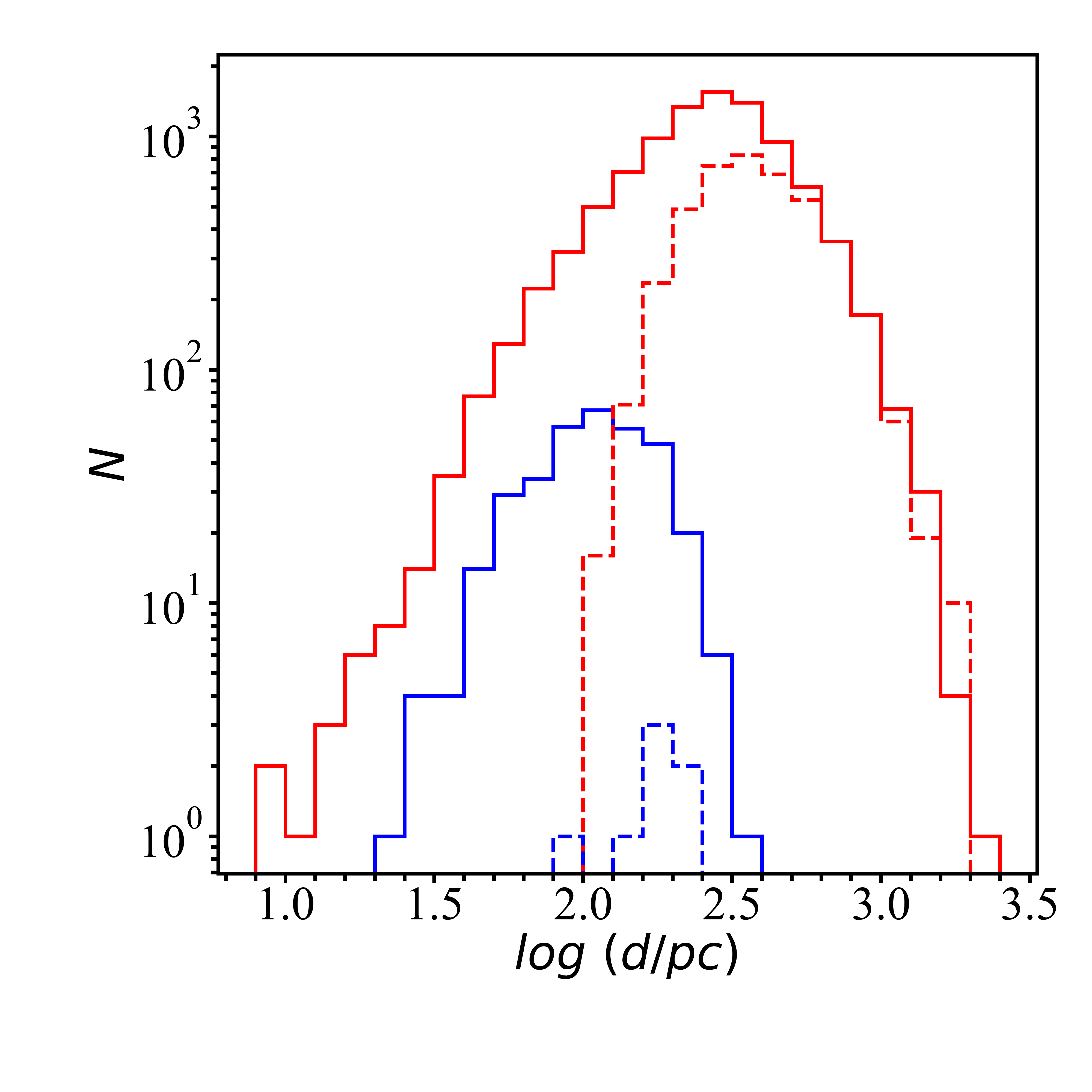}
\includegraphics[width=0.22\textwidth, trim=4.65cm 0.0cm 0.0cm 0.0cm, clip]{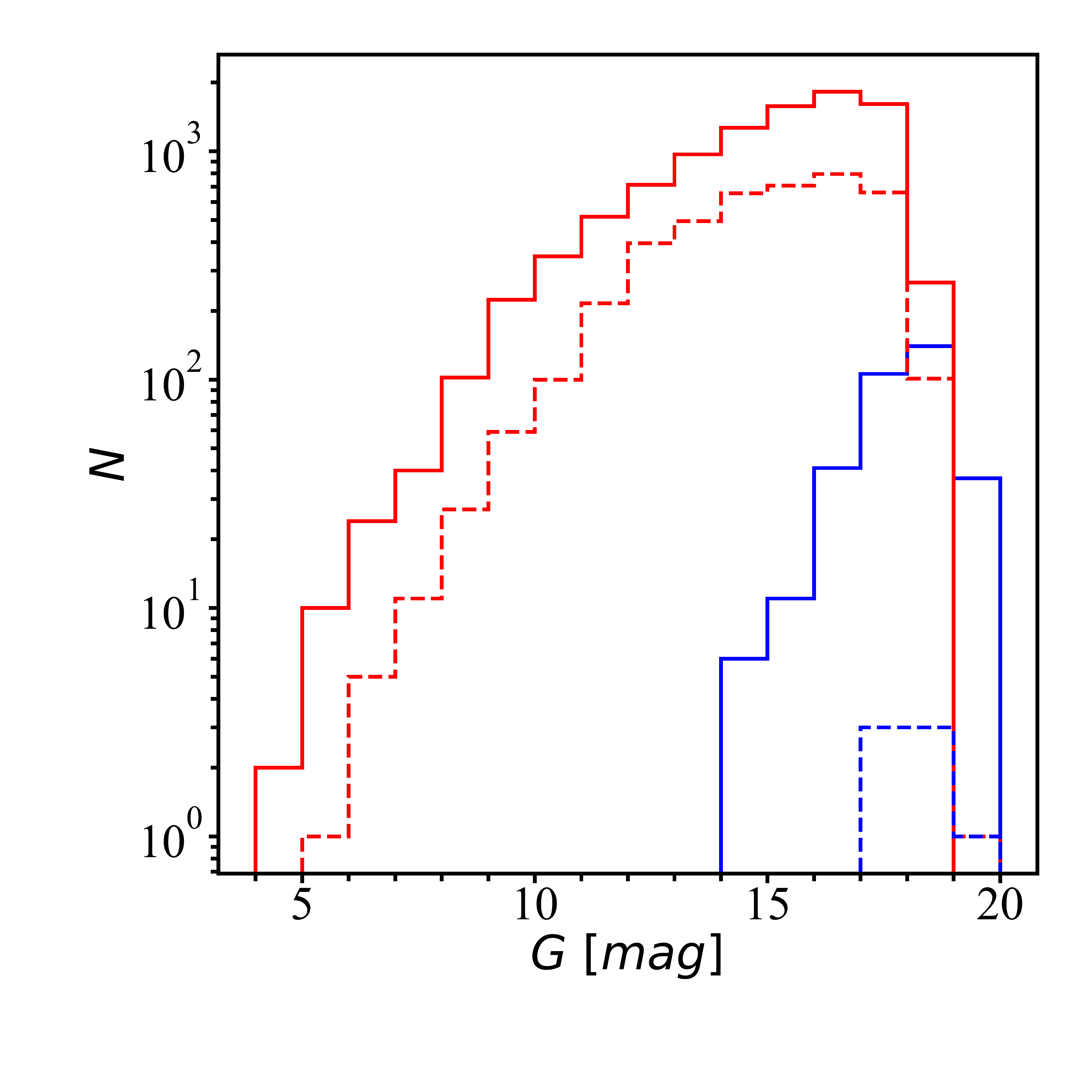}
\includegraphics[width=0.223\textwidth, trim=4.65cm 0.0cm 0.0cm 0.0cm, clip]{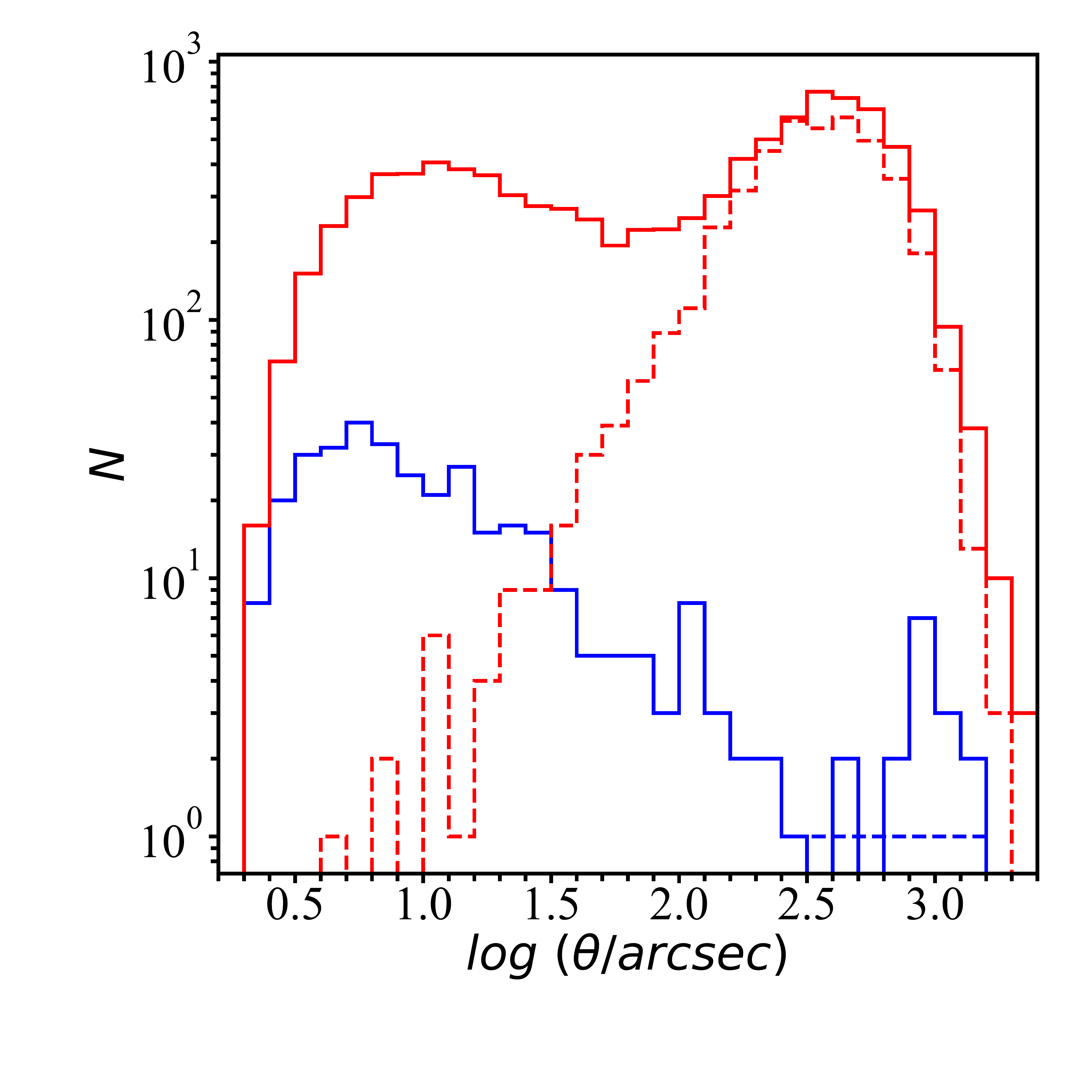}
\caption{Distributions of separations ($\log(s)$, first column), distances ($\log(d)$, second column), magnitude of the primary component (G, third column), and angular separation ($\log(\theta)$, fourth column) for the MS-WD (the solid red curves), WD-WD (the solid blue curves) binary candidates and the chance alignments (the dashed curves) selected from a shifted Gaia DR2 catalog. This catalog is produced by shifting all the objects of Gaia DR2 by 1\degr\ in both the right ascension and declination directions. Matching the original Gaia DR2 with the shifted catalog, the matched pairs are regarded as chance alignments if they pass our selection criteria for the MS-WD and WD-WD binaries.
}\label{fig:sample_distribution}
\end{figure*}

Figure \ref{fig:sample_distribution} shows the distributions of separations ($\log(s)$, the first column), distances ($\log(d)$, the second column), primary magnitude (G, the third column), and angular separation ($\log(\theta)$, the fourth column) for the MS-WD (the solid red curves), WD-WD (the solid blue curves) binary candidates and the chance alignments (the dashed curves) selected from a shifted Gaia DR2 catalog (see T20). For the MS-WD candidate sample, the distributions are dominated by contaminants (chance alignments) at large separations ($\theta>2.5$\,arcmin or $s>0.3$\,pc). For the WD-WD candidate sample, the impact of the contamination is relatively small. Around 10 chance alignments are mainly at $s>0.3$\,pc or $\theta>15$\,arcmin. The distances of the candidate binaries are less than 2.5\,kpc and 0.5\,kpc for the MS-WD and WD-WD data-sets (see the second sub-panel), respectively. The magnitudes in G-band are brighter than 19\,mag and 20\,mag for the primary components in the two samples (see the third sub-panel).

\begin{figure}[!t]
\centering
\includegraphics[width=0.45\textwidth, trim=0.0cm 3.3cm 0.0cm 0.0cm, clip]{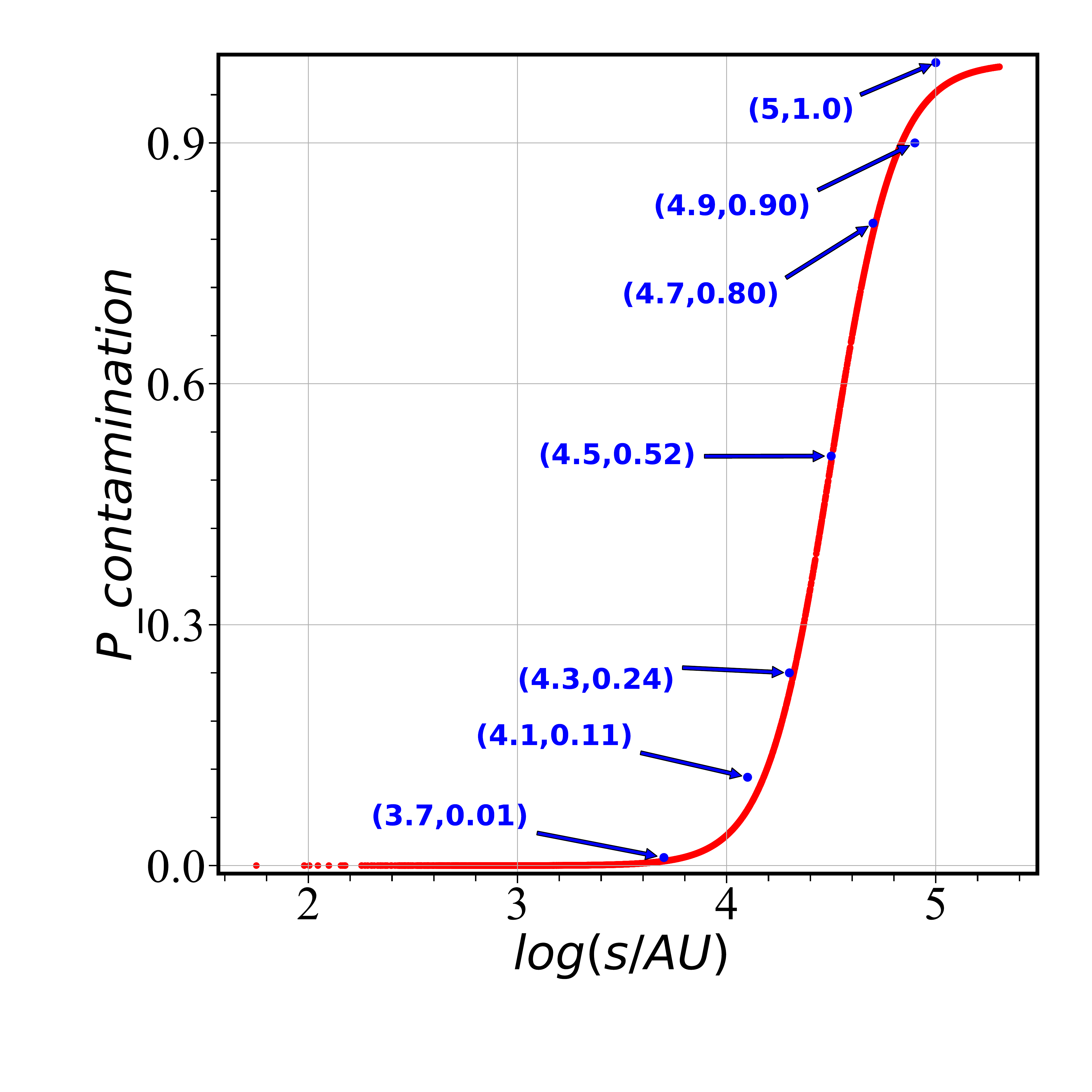}
\caption{The $\log(s/AU)$ vs. contamination rates for 9589 MS-WD candidates.
}\label{fig:contamination}
\end{figure}

We estimate the contamination rates of the samples with the approach adopted in T20. A ``shifted Gaia DR2 catalog''with coordinates ($\hat{\rm \alpha}$, $\hat{\rm \delta}$) has been constructed by shifting 1\degr\ from its original location (${\rm \alpha}$, ${\rm \delta}$), i.e., ($\hat{\rm \alpha}$, $\hat{\rm \delta}$) = (${\rm \alpha}$ + $\Delta\alpha^{*}/\cos(\delta)$, ${\rm \delta}$ + $\Delta\delta$), with $\Delta\alpha^{*}=\Delta\delta=1.0$\degr. We repeat the binary candidate identification procedure, now identifying pairs that pass the binary cuts when the coordinates of the ``primary'' are shifted relative to candidate secondaries. This procedure removes genuine binaries, but preserves chance alignment statistics. Using this method, we estimate the chance alignment rates in the different separation bins. Table \ref{tab:contaminate} specifies the contamination rates for the 9,589 MS-WD and 307 WD-WD samples in the different separation bins. Based on these contamination rates (assuming 100\% contamination rates at $\log(s)>5.0$), we obtain the contamination rates (P\_contamination) for each MS-WD system by linear interpolation. The smooth curve in Figure \ref{fig:contamination} demonstrates the best-fitted model of the contamination rates in $\log(s/AU)$ for 9,589 candidates. It is clear that the contamination rates sharply increase at $\log(s/AU)>4.0$. Note that for the contamination rates, we do not account for sky-position dependence in this present work. Given these contamination rates, we selected 4\,050 MS-WD pairs with contamination rates $<20\%$.

\begin{table}
 \begin{threeparttable}
\centering
\caption{Contamination rates of the initial candidates}
\label{tab:contaminate}
\begin{tabular}{p{2.1cm} | p{2.1cm}| p{2cm}} 
\hline
\hline
Separations&\multicolumn{2}{c}{Contamination Rates}\\
\hline
$\log(s/(AU))$  & MS-WD & WD-WD  \\
\hline
$(0.0,3.4]$ & $0\%$  & $0\%$  \\
$(3.4,4.0]$ & $<5\%$ &$0\%$  \\
$(4.0,4.2]$ & $\sim 10\%$ &$0\%$  \\
$(4.2,4.4]$ & $\sim 20\%$ & $0\%$  \\
$(4.4,4.6]$ & $\sim 50\%$ & $0\%$  \\
$(4.6,4.8]$ & $\sim 80\%$ & $\sim 100\%$  \\
$(4.8,5.0]$ & $\sim 90\%$ & $\sim 100\%$  \\
$(5.0,5.4]$ & $\sim 100\%$ & $\sim 100\%$  \\
\hline
\hline
\end{tabular}
 \end{threeparttable}
\end{table}

\section{Age Determination for WD components}
We derive the total system ages from the WDs, using the same approach as \citet{Fouesneau2019} adopted. In the pilot study, we employed a statistical technique \citep{Hipple_2006} implemented as a flexible software suite, BASE-9, which has demonstrated good performance on age estimation for individual field WDs \citep{Malley_2013, Webster_2015}. BASE-9 works by comparing a comprehensive stellar evolution model \citep{Hipple_2006} with observational data (the individual uncertainties are accounted for) in a wide range of possible passbands, e.g., the photometry in the four bands ($g_{\rm P1}$, $r_{\rm P1}$, $i_{\rm P1}$ , and $z_{\rm P1}$) of Pan-STARRS1 (PS1; \citet{Chambers_2011}) or (g,r,i and z) of Skymapper DR 1.1 \citep{Wolf_2019} and the three bands of {\it Gaia} are used in this study. Some ancillary information (e.g., parallax) and astrophysical knowledge are incorporated through the prior distributions. In the following sections, we describe how to set priors and build the comprehensive stellar evolution model for age constraints in detail. 

\subsection{The stellar evolution models for BASE-9}
BASE-9 integrates several theoretical models with various stages of stellar evolution, such as the models for MS and red giant stars \citep[e.g.][]{Girardi_2000, Dotter_2008, Yi_2001,Marigo_2017}, and for WD interior cooling \citep[e.g.][]{Althaus_1998, Montgomery_1999,Renedo_2010, Bischoff-Kim_2018}. Using the initial-final mass relation(IFMR) \citep[e.g.][]{Salaris_2009, Williams_2009}, the precursor and WD models are bridged into a comprehensive model which depicts stars in all of the main phases of stellar evolution \citep{Dyk_2009}. Given the WD evolution models and the WD atmosphere models \citep[e.g.][]{Bergeron_1995, Bergeron_2019}, the surface luminosities and temperatures are converted into magnitudes. In practice, we extract isochrones for MS stars from PAdova and TRieste Stellar Evolution Code \citep[PARSEC;][]{Marigo_2017}. The range of log(age) is from 7.42 to 10.13 with a increment of 0.01, and the range of [Fe/H] is from -2.0 to 0.5 with a increment of 0.5 as provided by PARSEC. We adopt the WD cooling model of \citet{Bischoff-Kim_2018}, the IFMR of \citet{Williams_2009}, and the WD atmosphere of \citet{Bergeron_2019}. Without a spectrum, we cannot determine the atmospheric type (DA or DB) of WDs, but \citet{Tremblay_2008} found that only about 25\% of WDs have helium-dominated atmospheres. It is therefore a safe assumption that the majority of WDs whose posterior distance probabilities are consistent with their candidates MS companion parallaxes are indeed DAs.

\subsection{Priors on the five inferred parameters}
BASE-9 is the abbreviation of Bayesian Analysis for Stellar Evolution with Nine Parameters. This means that BASE-9 supports up to nine free parameters, which are the age, metallicity, distance modulus, line-of-sight absorption, helium abundance, and parameters of the IFMR for a cluster, and the primary mass, secondary mass (if a binary), and cluster membership probability for each star. In this study, we fix helium abundance at the scaled solar value within the PARSEC models. All WDs are considered as single stars only, because we assume that the WD in the common proper motion binary is not itself an unresolved double. Additionally, because we are analyzing these stars individually, the parameter for cluster membership is meaningless. Thus for each star BASE-9 is fitting five parameters: age, [Fe/H], distance modulus, $A_V$, and the ZAMS mass of the WD. Choosing reasonable priors is conducive to fast convergence of the free parameters and obtaining reliable results in BASE-9.

For the prior of line-of-sight extinction $A_V$, we use the values as described in Section \ref{sect:data}, and set the dispersion $\sigma$($A_V$)=1/3 $A_V$ (or 0.03 mags if $A_V$=0). For the distance modulus $\left(m-M\right)_0$ prior, we derive the values directly from parallax by:
\begin{equation}
\label{eq:Distmod} 
\left(m-M\right)_0=-5\log\left(\varpi/mas\right)+10,
\end{equation}
\noindent where $\varpi$ is the parallax from {\it Gaia}. The value $ \left(m-M\right)_0$ is the intrinsic distance modulus, yet BASE-9 works with the observed modulus, so we use $(m-M)_V$ as the input by considering the extinction:
\begin{equation}
\label{eq:Distmod_v} 
\left(m-M\right)_V=\left(m-M\right)_0+A_V.
\end{equation} 
\noindent The dispersion $\sigma$(m-M) is given by
\begin{equation}
\label{eq:eDistmod}
\sigma\left(m-M\right)=5\sigma\left(\varpi\right)/\left(\varpi\ln10\right),
\end{equation} 
\noindent where $\sigma$($\varpi$) is the uncertainty of parallax from {\it Gaia}. 
For the age $\tau$, we set flat prior in $\log(\tau)$ as $\langle\log(\tau/\rm yr)\rangle = 9.5$, and $\sigma(\log(\tau/\rm yr)) = \inf$, since the average age is around 3\,Gyr for the majority of WDs. 

Due to the fact that we do not yet have spectroscopic abundances (of the MS primary), we empirically set the prior distribution on metallicity to be a broad Gaussian with a mean $\langle\rm[Fe/H]\rangle = -0.5$\,dex, and $\sigma$([Fe/H]) = 1.0\,dex, which covers a wide range of metallicity to match all the isochrones from PARSEC which allows [Fe/H] from -2.0 to 0.5. 

Because there is no information on [Fe/H] contained in the WD's properties, the posterior [Fe/H] distribution is essentially equal to its prior distribution. Thus actually only four parameters are fitted for each WDs. Note that BASE-9 can provide meaningful constraints on all four parameters for an individual WD, even though the number of data points (parallax and photometric bandpasses) may be lower. For instance, a cool, high mass WD must have evolved from a short-lived high mass precursor star and its total age will therefore be dominated by the length of time it has been cooling as a WD. Such a star could deliver a precise age from just two or three band photometry along with a precise parallax. Nonetheless, more bandpasses are preferable to help constrain $A_V$ and to decrease the impact of photometric uncertainty in any particular bandpass.

\subsection{Age Constraints}
\label{sect:7_3filters}

Based on the above models and priors, we ran BASE-9 on each cWD individually. BASE-9 uses a Monte Carlo-Markov chain (MCMC) algorithm to adjust stellar physical parameters at each iteration. BASE-9 allows us to use observed photometric magnitudes to fit the models with the parameters including age, ZAMS mass, [Fe/H], $\rm (m-M)_V$, and $A_V$. {\it Gaia} provides photometry for each cWD in three bands, i.e., $\rm G$, $\rm G_{BP}$, $\rm G_{RP}$. To obtain more photometric information, we cross-match the sample with PS1 PV3 \citep{Chambers_2011} and Skymapper DR 1.1 \citep{Wolf_2019}, and obtain the photometry in the $g_{\rm P1}$, $r_{\rm P1}$, $i_{\rm P1}$, and $z_{\rm P1}$ bands for 3603 cWDs (3109 from MS-WD, 494 from WD-WD) and griz bands for 1202 cWDs (1006 from MS-WD and 196 from WD-WD). Among the cross-matched sources, 711 cWDs (556 from MS-WD and 109 from WD-WD) were observed by both the PS1 and the Skymapper. For this part of cWDs, we only use the photometry from PS1 and {\it Gaia}. For 524 cWDs (491 from MS-WD and 33 from WD-WD) which are observed neither by PS1 nor by Skymapper, we only use the photometry of {\it Gaia} passbands to constrain the physical parameters.

\begin{figure*}
\centering
\includegraphics[width=0.49\textwidth, trim=0.0cm 0.0cm 0.0cm 0.0cm, clip]{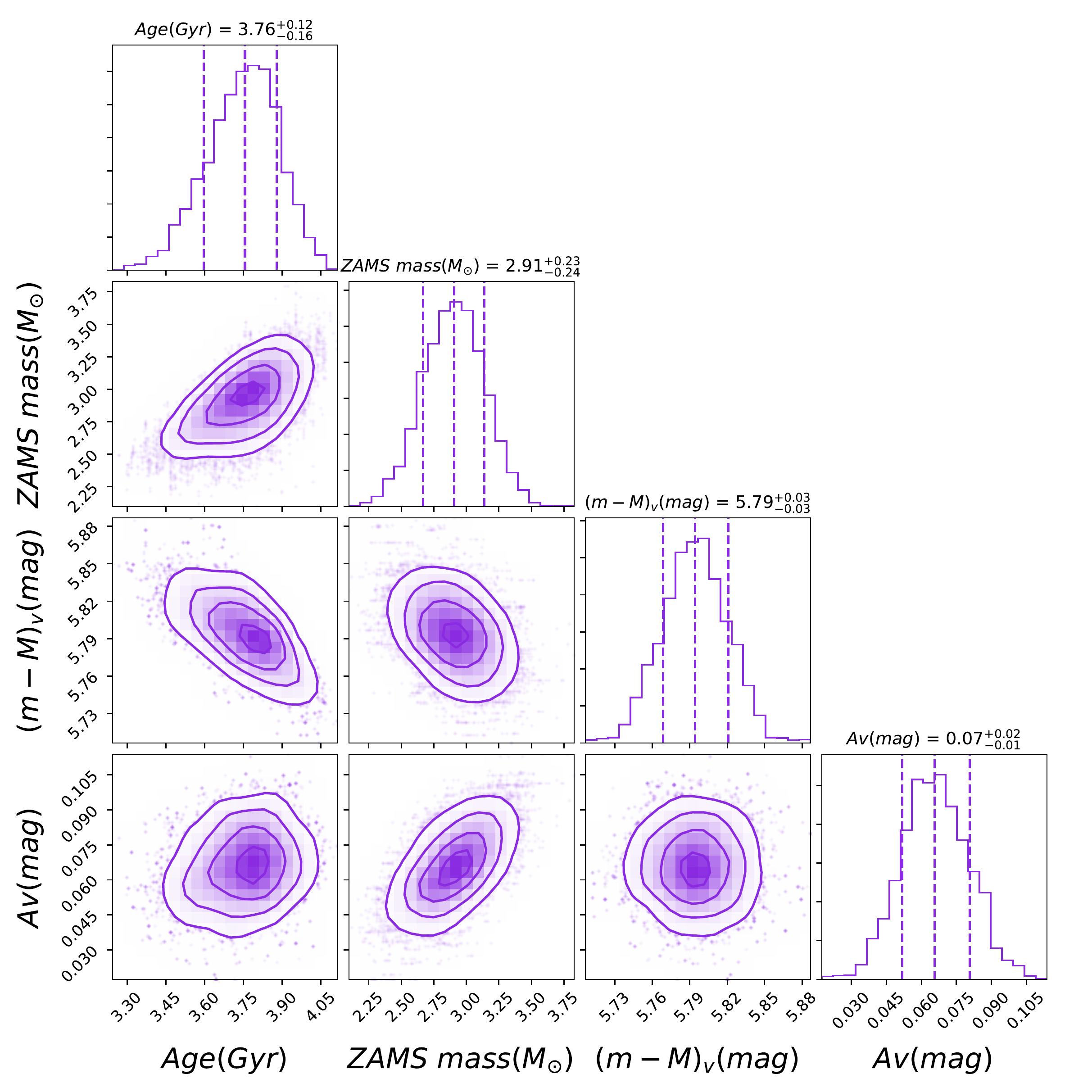}
\includegraphics[width=0.49\textwidth, trim=0.0cm 0.0cm 0.0cm 0.0cm, clip]{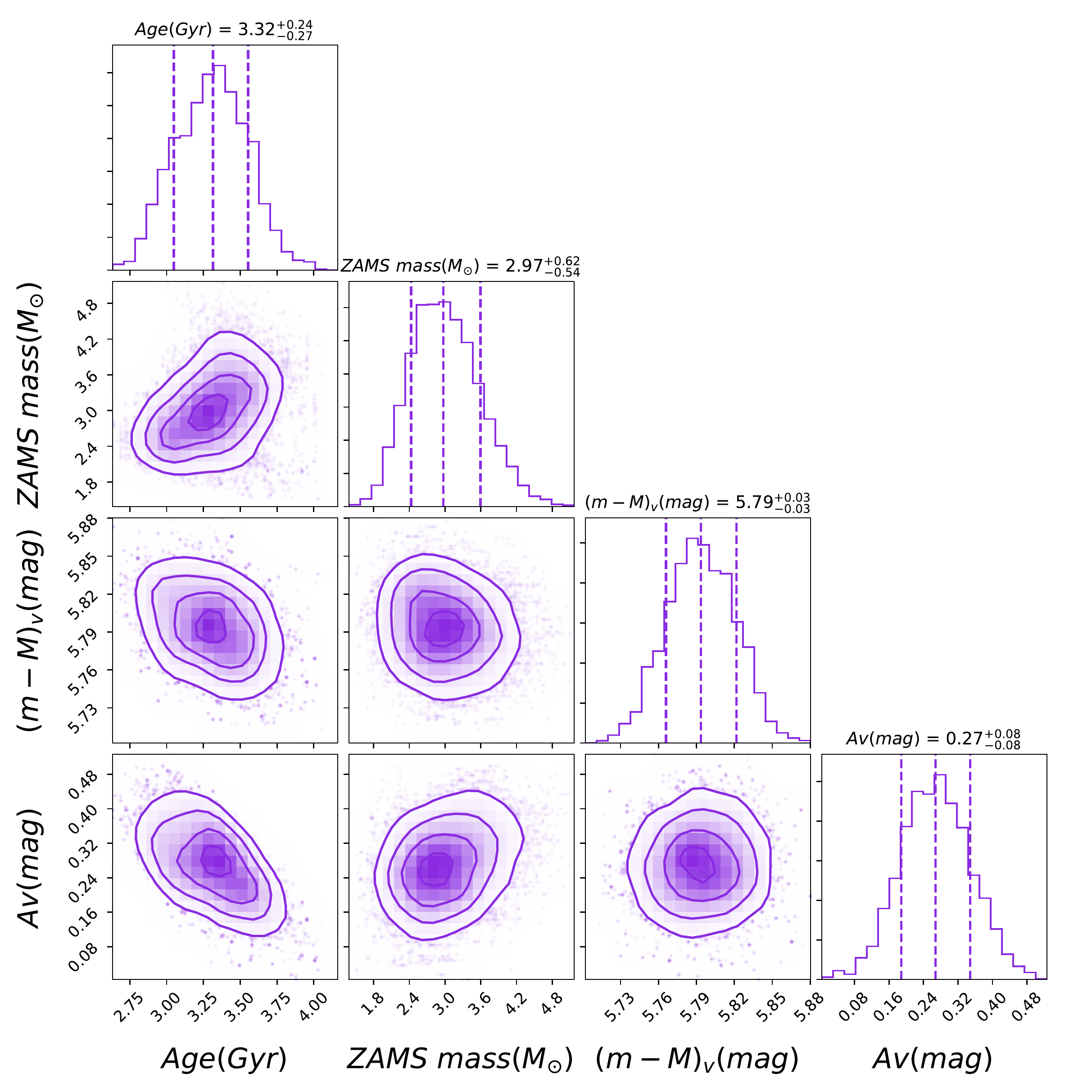}
\caption{The contours and marginalized (the 16$^{th}$ and 84$^{th}$ percentiles) probability distributions of the four parameters from the MCMC sampling for one typical cWD (source\_id = 2759526902977240192), based on two sets of photometric bands. Left: the case with seven bands ($g_{\rm P1}$, $r_{\rm P1}$, $i_{\rm P1}$, $z_{\rm P1}$, and $\rm G$, $\rm G_{BP}$, $\rm G_{RP}$); Right: the same cWD with only three bands from {\it Gaia}. The four parameters are Age\,(Gyr), ZAMS mass\,($M_{\odot}$), {\bf apparent distance moduli\,(mag)} and Av\,(mag). The ages are matched well in the two cases, but the additional photometry from the four PS1 bands significantly improve the precision of the age. The case with the seven bands (the left panel) has a smaller ZAMS mass than the case with only three {\it Gaia} bands (the right panel). The apparent distance modulus are consistent with each other in two cases, and Av precision of the former is better than the latter.}
\label{fig:posterior dstribution}
\end{figure*}  

Figure \ref{fig:posterior dstribution} illustrates the contours and marginalized probability distributions of the four parameters from the MCMC sampling for one typical cWD. The left panel displays the constraint of the cWD with the photometry from the seven bands (four from PS1 or Skymapper DR1.1, and three from Gaia), while the right panel presents the case on the same cWD, but with the photometry only from the three Gaia bands. The final ages are matched with each other, but the age precision of the former (3.76$^{+0.12}_{-0.16}$\,Gyr) is significantly better than the latter (3.32$^{+0.24}_{-0.27}$\,Gyr). It is reasonable since more observational information usually gives a tighter constraint on one parameter. The ZAMS masses precision of the former (2.91$^{+0.23}_{-0.24}\rm\,M_{\odot}$) are also better than the latter (2.97$^{+0.62}_{-0.54}\rm\,M_{\odot}$), and they are matched with each other with their posterior distributions overlapping significantly. To demonstrate how the procedure constrains the extinction, we use the same prior extinction value in this experiment, i.e., 0.26\,mag obtained from SFD, which puts an upper bound on our estimate for both the cases. We found that the posterior distributions are broad in the case of three bands, nearly identical to the prior suggesting that little knowledge was gained. In contrast, seven band analysis leads us to narrow extinction posterior distributions. Moreover, we cross-validate our extinctions with the 3D map of \citet{Green_2019} (the extinction is 0.0\,mag for this WD), and we find our results agree better when using seven bands than three bands. It indicates that the additional photometry from the four PS1 bands or Skymapper can effectively constrain the extinction relative to the case with only the three Gaia bands. The posterior apparent distance moduli are consistent with each other in the two cases, both of them remain nearly the same as the prior (i.e, 5.53+0.26\,mag, where the 5.53\,mag is derived from Gaia parallax according to Equation \ref{eq:Distmod}, and the 0.26\,mag is the extinction value obtained from SFD). It suggests that our procedure seems not to be able to well constrain the apparent distance modulus for both cases. Furthermore if the prior apparent distance moduli are biased by the extinction, so too are the posterior values. Fortunately, the vast majority of extinction values have been obtained for our cWDs from the 3D map of \citet{Green_2019}, the exact extinction will not bias the apparent distance modulus.

\section{Results and performance}

We are almost exclusively interested in the data subset, where the (binary) contamination rates are lower than $20\%$. Therefore, we focus on determining the ages and other parameters for the 4050 cWDs from this dataset with BASE-9. We discard the cases with very poor age convergence (probably caused by the large photometric errors) in the MCMC sampling, leaving us with estimates for 3551 MS-WD candidate binaries.

\subsection{Ages and other parameters}
\label{sect:parameter statistical}

Among the four free parameters, the distance modulus is actually a direct observable, constrained by the trigonometric parallax. We use the more precise parallax among the two components as the prior mean value. In most cases, the brighter MS component has a more precise parallax. 

\begin{figure*}
\includegraphics[width=0.3\textwidth, trim=0.45cm 0.2cm 0.2cm 0.2cm, clip]{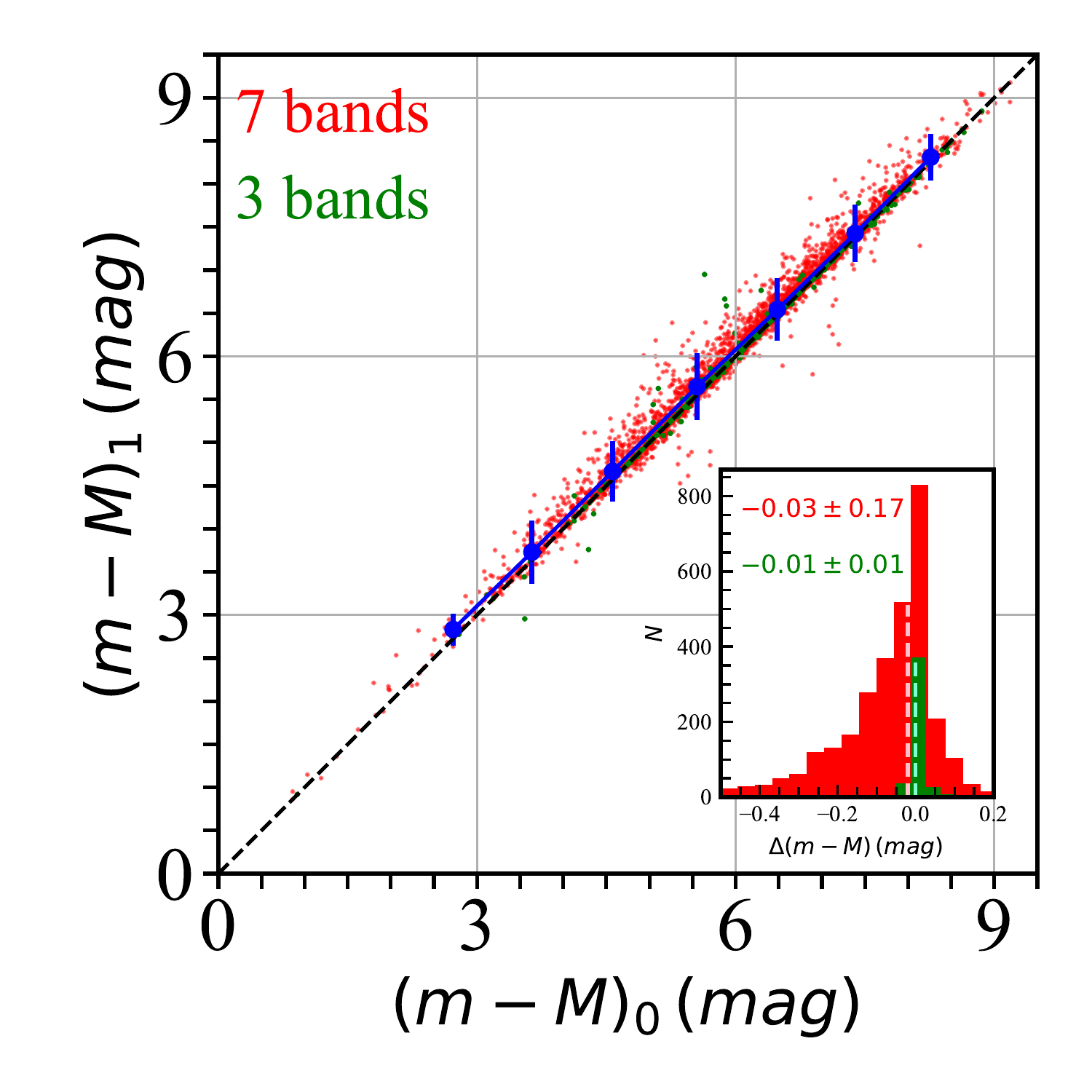}
\includegraphics[width=0.305\textwidth, trim=0.3cm 0.2cm 0.2cm 0.2cm, clip]{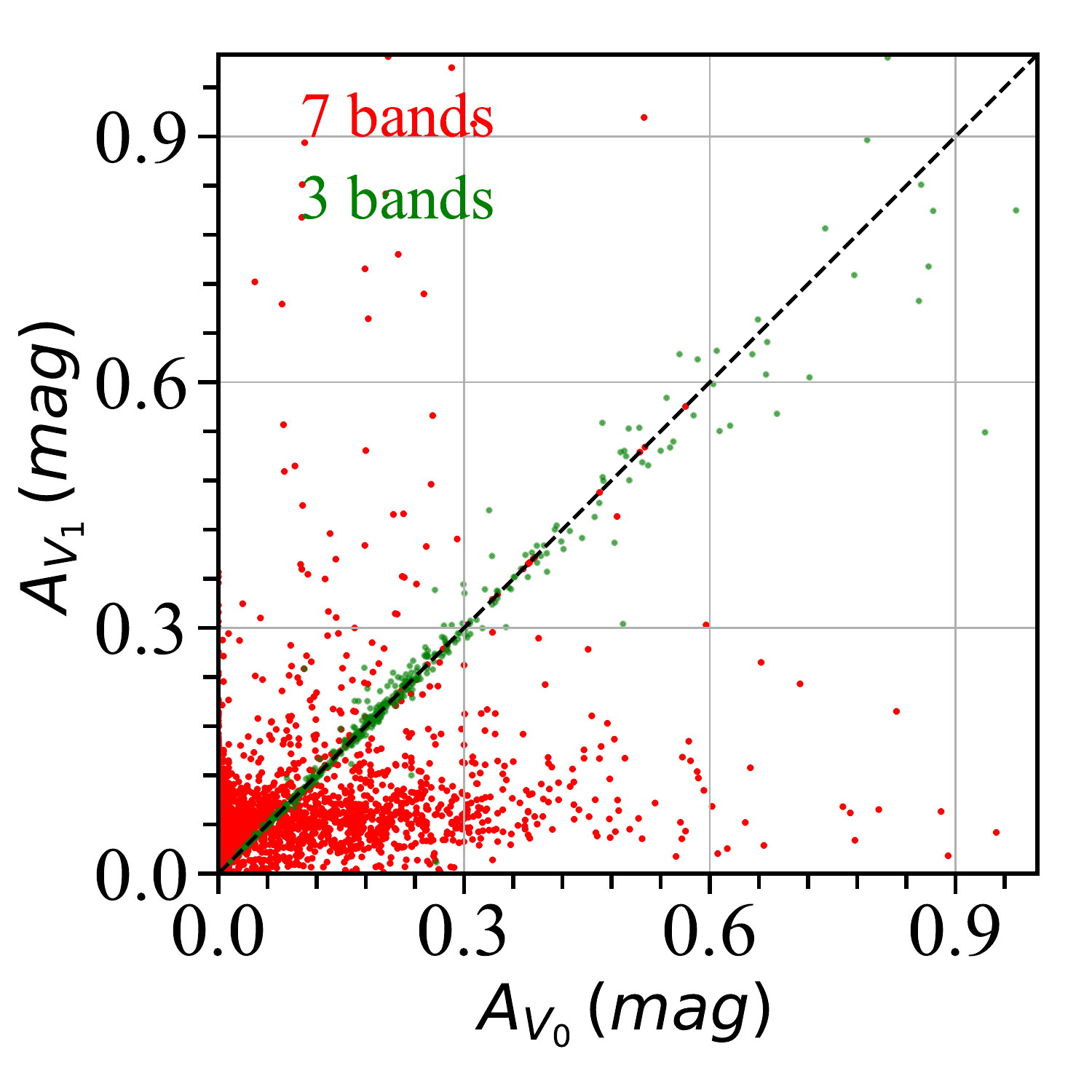}
\includegraphics[width=0.3\textwidth, trim=0.5cm 0.2cm 0.2cm 0.2cm, clip]{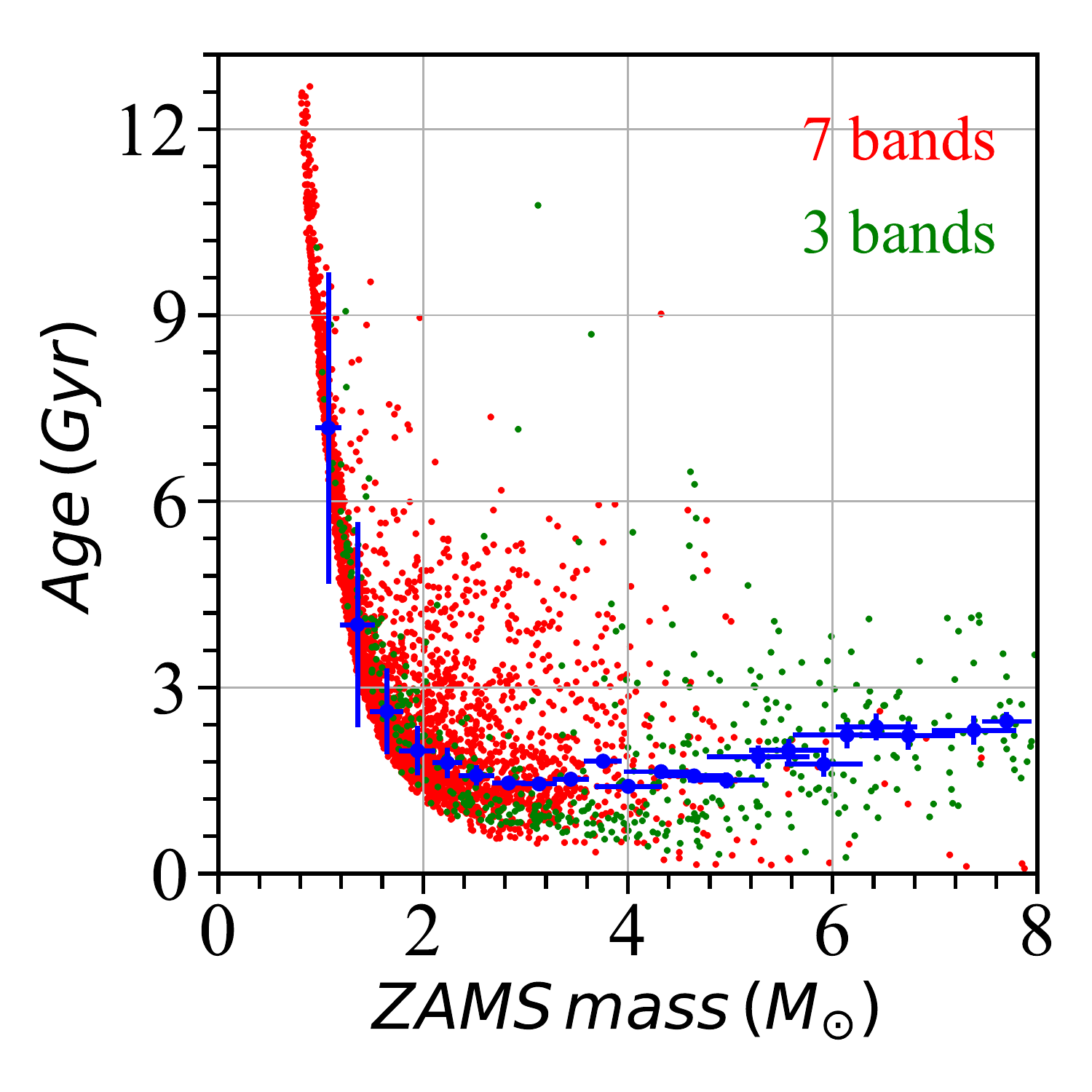}
\caption{Left: Comparison between the prior mean ($(m-M)_0$, see Equation \ref{eq:Distmod}) and posterior mean ($(m-M)_1$) distance moduli for 3551 WDs, the $(m-M)_1$ is equal to $(m-M)_V-A_{V1}$, where $(m-M)_V$ is the posterior apparent distance modulus and $\rm A_{V_1}$ is the posterior extinction. The blue dots and bars indicate the median and {\it rms} of $(m-M)_1$ in the different $(m-M)_0$ bins. The insert sub-panel displays the histogram of $\Delta\rm(m-M)$ ($ \equiv \rm (m-M)_{0} - (m-M)_1$) for cWDs estimated from seven bands (red) and three bands (green). The two dashed lines mark the location of $\left<\Delta\rm(m-M)\right>$ for these two types of WDs in the sub-panel. Middle: Comparison between the prior mean $A_{V_0}$ and posterior $A_{V_1}$ extinctions for 3551 cWDs. Right: Age vs. ZAMS mass for 3551 cWDs. The blue dots and bars represent the median of ages and median of age uncertainties in the different ZAMS mass bins. The dashed black lines are the one-to-one line in the left and middle panels.}
\label{fig:Distmod-compare}
\end{figure*}

 The left panel of Figure \ref{fig:Distmod-compare} displays the comparison of the two distance moduli, i.e., the prior $\rm (m-M)_{0}$, which is directly derived from Gaia parallax according to Equation \ref{eq:Distmod}, and the posterior $\rm (m-M)_1$, which is equal to $\rm (m-M)_V-A_{V_1}$, where $\rm (m-M)_V$ is the posterior apparent distance modulus and $\rm A_{V_1}$ is the posterior  extinction. The black dashed line is the one-to-one line. The blue dots and bars indicate the median and root-mean-square ({\it rms}) of the $\rm (m-M)_1$ in the different $\rm (m-M)_0$ bins for 3551 cWDs. The MCMC fitting suggests that the majority of cWDs are {\it bona fide} WDs that are well fit by the models. Note the values of $\rm (m-M)_1$ for 3076 cWDs fitted by seven bands (red) and 475 cWDs fitted by three bands (green) are sightly larger than that of $\rm (m-M)_0$ for most distance modulus bins. This can be found from the histogram distribution of $\Delta\rm(m-M)$ ($ \equiv \rm (m-M)_{0} - (m-M)_1$) in the insert sub-panel. The median of $\Delta\rm(m-M)$ for cWDs estimated by seven bands (red) is -0.03$\pm$0.17\,mag, and the median of $\Delta\rm(m-M)$ for cWDs estimated by three bands (green) is -0.01$\pm$0.01\,mag. This suggests that the two types of distance moduli are matched within 1$\sigma$. It is worth mentioning that there are more cWDs of which the $\Delta\rm (m - M)$ are negative in the case of seven bands (red) relative to the case of three bands (green), as shown by the inset histograms in the left sub-panel. The main reason for this is that the posterior apparent distance moduli $(\rm(m - M)_V)$ are likely over-estimated by the extinction for a small part of cWDs of which the extinction values are obtained from SFD, yet the posterior extinction $\rm A_{V_1}$ is close to the smaller true value in the case of seven bands, as discussed in Section 3.3. This effect will give rise to an over-estimated $(m - M)_1$, and subse- quently lead to negative values of $\Delta\rm(m-M)$ for many cWDs.

The extinction ($A_V$) is another observable, which can be quickly obtained from the known extinction map. In this work, we obtained the value of $A_V$ for each cWD in Section \ref{sect:data} as a prior value in the BASE-9 MCMC fitting. By comparing the posterior $A_V$ (i.e., $A_{V_1}$) with the prior $A_V$ (i.e., $A_{V_0}$) for 3551 cWDs, we found that the prior values are systematically larger than the posterior ones, as shown in the middle panel of Figure \ref{fig:Distmod-compare}. This suggests that the prior $A_V$ are probably over-estimated for most of cWDs. In the case of only three bands (green), the $A_{V_0}$ values are consistent with the posterior ones. This demonstrates that the photometry in only three bandpasses can not well constrain $A_V$.

The initial mass of a star always determines its lifetime and influences its location on the CMD. So, here we discuss the two parameters (i.e., age and ZAMS mass), simultaneously. The greater the initial mass, the shorter its life as a star, and the more massive (and hence less luminous) its WD descendant. Therefore, for larger initial stellar masses of a star we should expect younger system lifetimes, as illustrated in the right-hand panel of Figure \ref{fig:Distmod-compare}. The plot displays the distribution of system ages vs. ZAMS mass for 3551 cWDs from MS-WD binaries. The blue dots and bars represent the median of ages and median of age uncertainties in different ZAMS mass bins. This Figure also indicates that the age uncertainties drop rapidly with ZAMS mass $>$ 1.8 \Msun. The age uncertainties of low mass stars are large because of the increasing fraction of the system lifetime spent on the MS (rather than as a cooling WD). The results from the seven and three bands are demonstrated with the red and green dots, respectively. It can be shown that there are no WDs descending from stars with ZAMS mass $<$ 0.8 \Msun, due to the finite age of the Galaxy. The main reason that there are no WDs in the upper right part is that they've been WDs for too long and have therefore cooled below Gaia's magnitude-limited detection threshold. The majority of our WDs have ages of 1$\sim$6\,Gyr, and the masses of most WDs in a range of 1$\sim$4 \Msun. This approach is well-suited to obtain precise ages for low-mass stars.

Figure \ref{fig:uncertaines-age-mass} illustrates the distribution of prior distance and relative age uncertainties for 3551 cWDs (left). The relative age uncertainties of 3076 cWDs (red dots) estimated from seven bands and 475 cWDs (green dots) estimated from three {\it Gaia} bands. The red (green) dots and bars represent the median and uncertainty values of the relative age error in different distance bins for the WDs with seven (three) bands. It can be concluded that the relative age uncertainties of WDs tends to increase with distance, and the relative age uncertainties of WDs estimated from three bands is larger than those estimated from seven bands. The right panel is the histogram distribution of the relative age uncertainties for 3551 cWDs (black) from the primary dataset and 3076 cWDs (red) estimated from seven bands, respectively. As the figure shows, there are 903, 765 ($\sim$25\%, $\sim$25\%) cWDs of which the relative age uncertainties are smaller than 10\%, 665, 583  ($\sim$19\%, $\sim$19\%) cWDs of which the relative age uncertainties are between 10\% and 20\%, 655, 608 ($\sim$19\%, $\sim$20\%) cWDs of which the relative age uncertainties are between 20\% and 30\%, 894, 829 ($\sim$25\%, $\sim$27\%) cWDs of which the relative age uncertainties are between 30\% and 50\%, and the remaining 434, 291 ($\sim$12\%, $\sim$9\%) of which the relative age uncertainties are larger than 50\%. These two panels indicate that the distance (i.e., parallax) and the number of bands are the two key factors that affect the age uncertainties. High precision {\it Gaia} parallaxes tightly constrain the present mass of cool WDs. Yet, when the mass is mapped back onto the ZAMS mass, the uncertainties in mass can be amplified to large uncertainties in ages. In addition, the IFMR is not known perfectly, and therefore a small change in the model also might amplify the uncertainty in precursor mass and the pre-WD ages, particularly for WDs with ZAMS $<2\rm\, M_{\odot}$. Thus, for those objects, we can derive a precise cooling age, but not a precise total age.
Under an assumption that the components are co-eval in a binary system, it is statistically safe to assign a WD age to its companion MS star for these binaries with contamination rates $<20\%$. Table \ref{tab:s<4.4} describes the catalog which includes the ages and other parameters of the 3551 field MS components. This catalog will be released on-line.

We cannot determine [Fe/H] for a WD based on the photometric information and instead constrain the progenitor's metallicity based on the companion MS star's metallicity. The IFMR is also metallicity independent. BASE-9 is therefore unable to extract any new information on the metallicity from these data and as a result the posterior [Fe/H] distribution is essentially equal to its prior distribution. 

\begin{figure}
\centering
\includegraphics[width=0.5\textwidth, trim=2.5cm 0.0cm 0.0cm 0.0cm, clip]{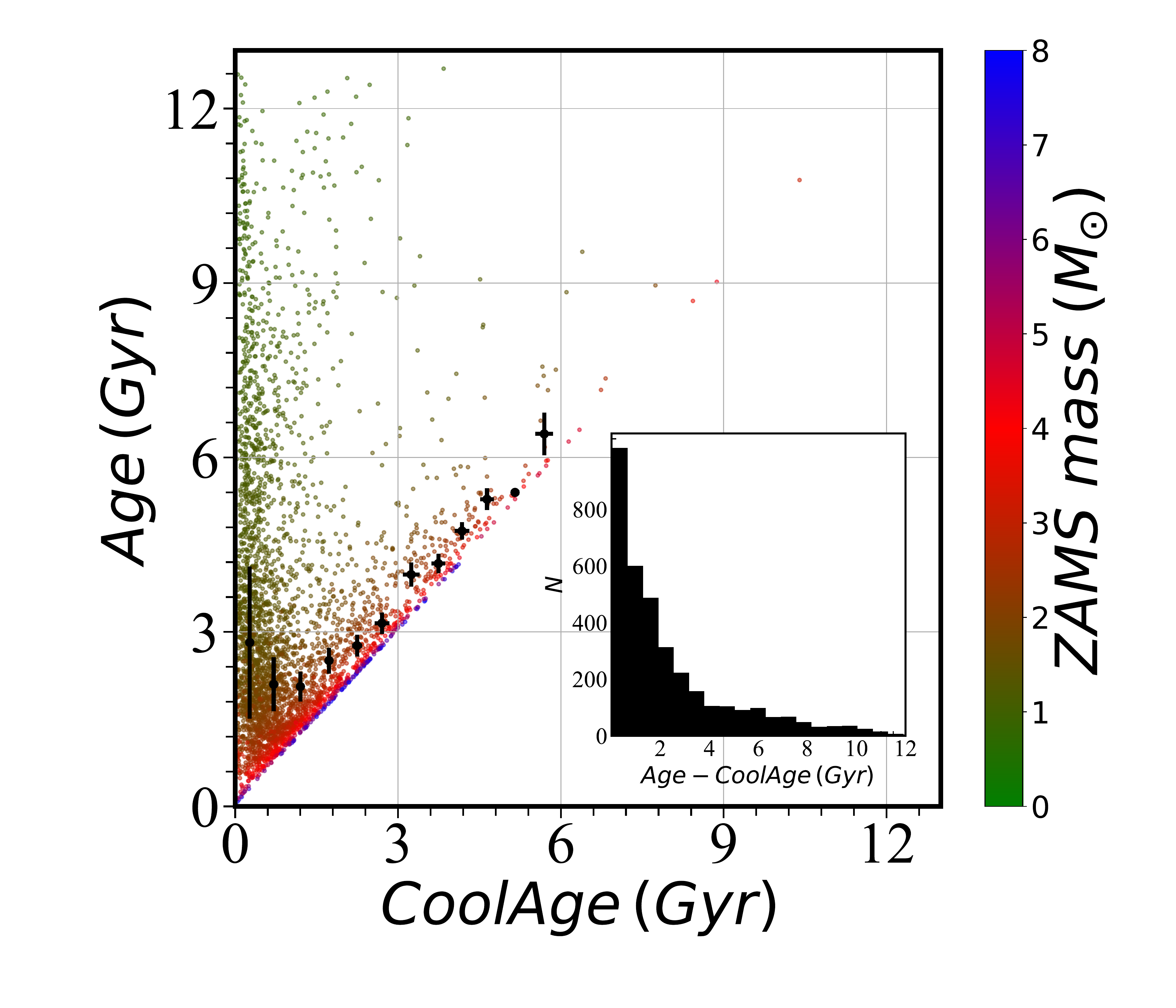}
\caption{The scatter distribution of the total age vs. cooling age of the 3551 cWDs, with points color-coded by ZAMS mass. The black points and error bars represent the median of total ages and the median uncertainties of ages in the different bins of cooling ages. The inset panel illustrates the histogram of the differences between the total ages and cooling ages.}
\label{fig:cooling-total-age}
\end{figure}

\begin{figure}
 \includegraphics[width=0.238\textwidth, trim=0.0cm 0.0cm 0.0cm 0.0cm, clip]{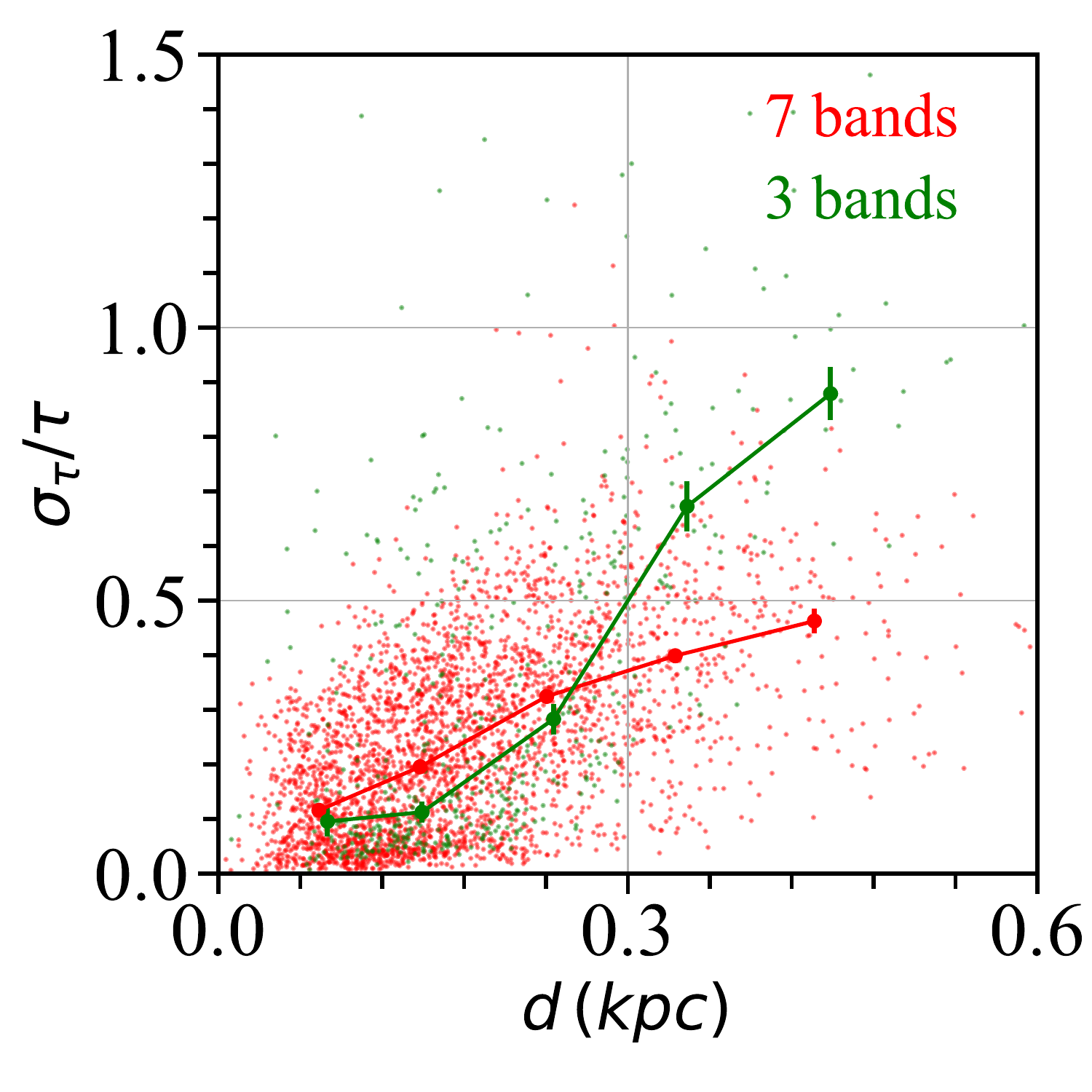}
 \includegraphics[width=0.23\textwidth, trim=1.6cm 0.2cm 0.0cm 0.0cm, clip]{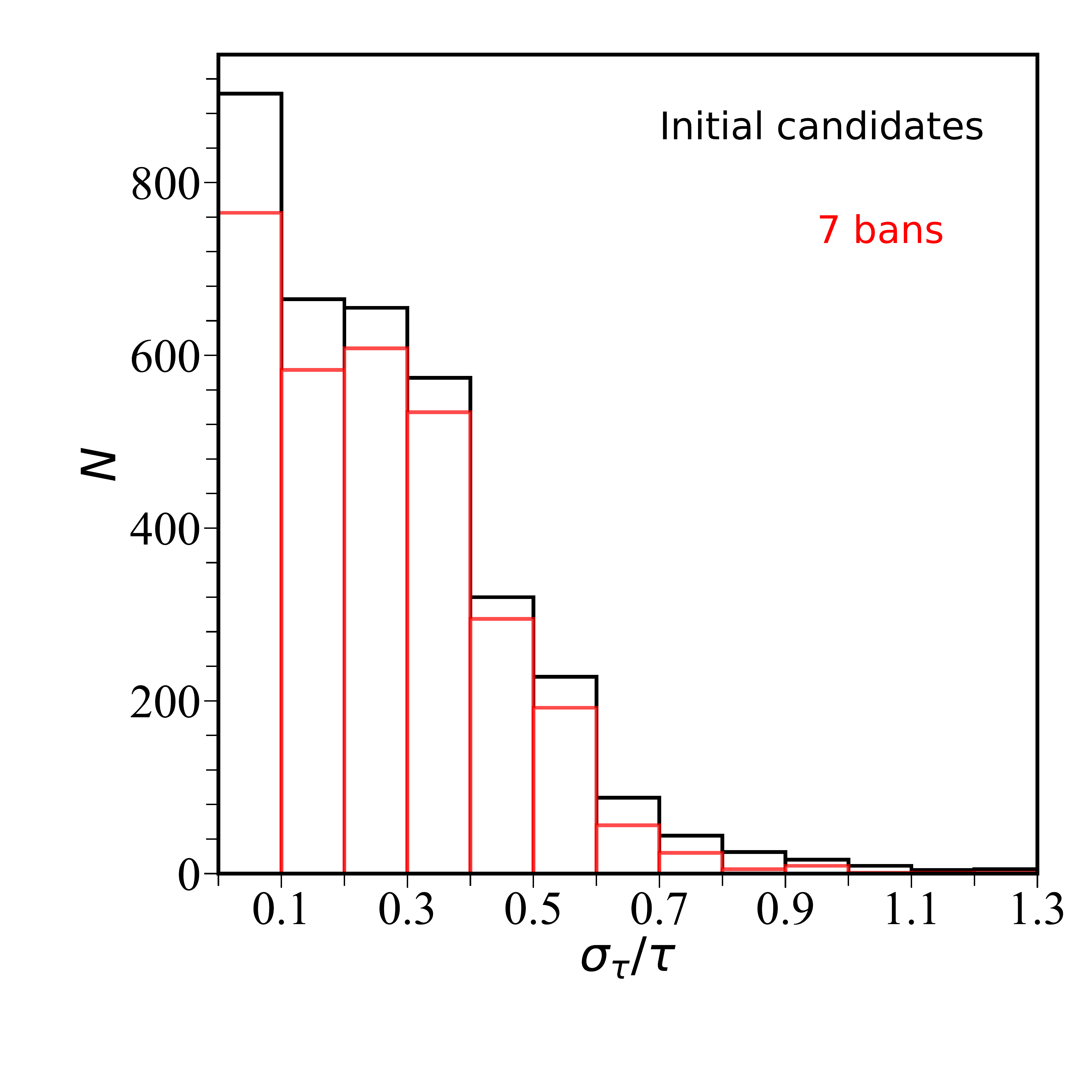}
\caption{Left: Distance vs. relative age uncertainties for 3551 cWDs. The red dots and bars represent the median and uncertainty values of the relative age errors for the WDs with seven bands, and the green dots and bars represent the median and uncertainty values of the relative age errors for the WDs with three bands. Right: The histogram distribution of the relative age uncertainties for cWDs with contamination rate $<20\%$ (black, 3551 cWDs), and a sub-sample of WDs estimated with seven bands (red, 3076 cWDs). The comparison indicates that the distance (i.e., parallax) and the number of bands are the two key factors that affect the age uncertainties.
}
\label{fig:uncertaines-age-mass}
\end{figure}

Figure \ref{fig:cooling-total-age} displays a comparison of the total age (y-axis) and the cooling age (x-axis) of the 3551 cWDs, with points color-coded by ZAMS mass. The black points and error bars represent the median of total ages and the median uncertainties of ages in the different bins of cooling ages. This diagram shows that there is not a typical case for WD cooling age vs. total age. There are a large number of cWDs that have been cooling for less than 1\,Gyr yet have total ages from a few to more than 12\,Gyrs. These cWDs are relatively easy to find because they are hot and bright, yet their range of precursor masses (see color-indicated masses in sidebar) means their MS lifetimes varied tremendously. There is also a distribution of cWDs with total ages not much greater than their cooling ages (the diagonal distribution dominated by red points), which are the result of higher mass stars that evolved more quickly on the main sequence. The inset panel illustrates the histogram of the difference between the total ages and cooling ages.

\subsection{Age validation} 
\label{Age validation}

We use two approaches to validate the WD ages derived from BASE-9. One method is by means of some well known clusters, the other is to use the WD-WD catalog for cross-validation. 

In the first approach, we choose several clusters whose ages have been determined precisely in the literature, such as Praesepe, NGC 6674, Comaber, and NGC 2682 \citep{Collaboration_2018}. In the four clusters, we find 6, 3, 1, and 1 WDs, respectively. We used our approach to determine the ages of these WDs. Table \ref{tab:Age} specifies the ages ($\log(\tau)$) from BASE-9 (the fourth column) and the literature (the last column). The two types of ages are highly consistent with each other.

\begin{figure}
\centering
\includegraphics[width=0.5\textwidth, trim=2.5cm 0.0cm 0.0cm 0.0cm, clip]{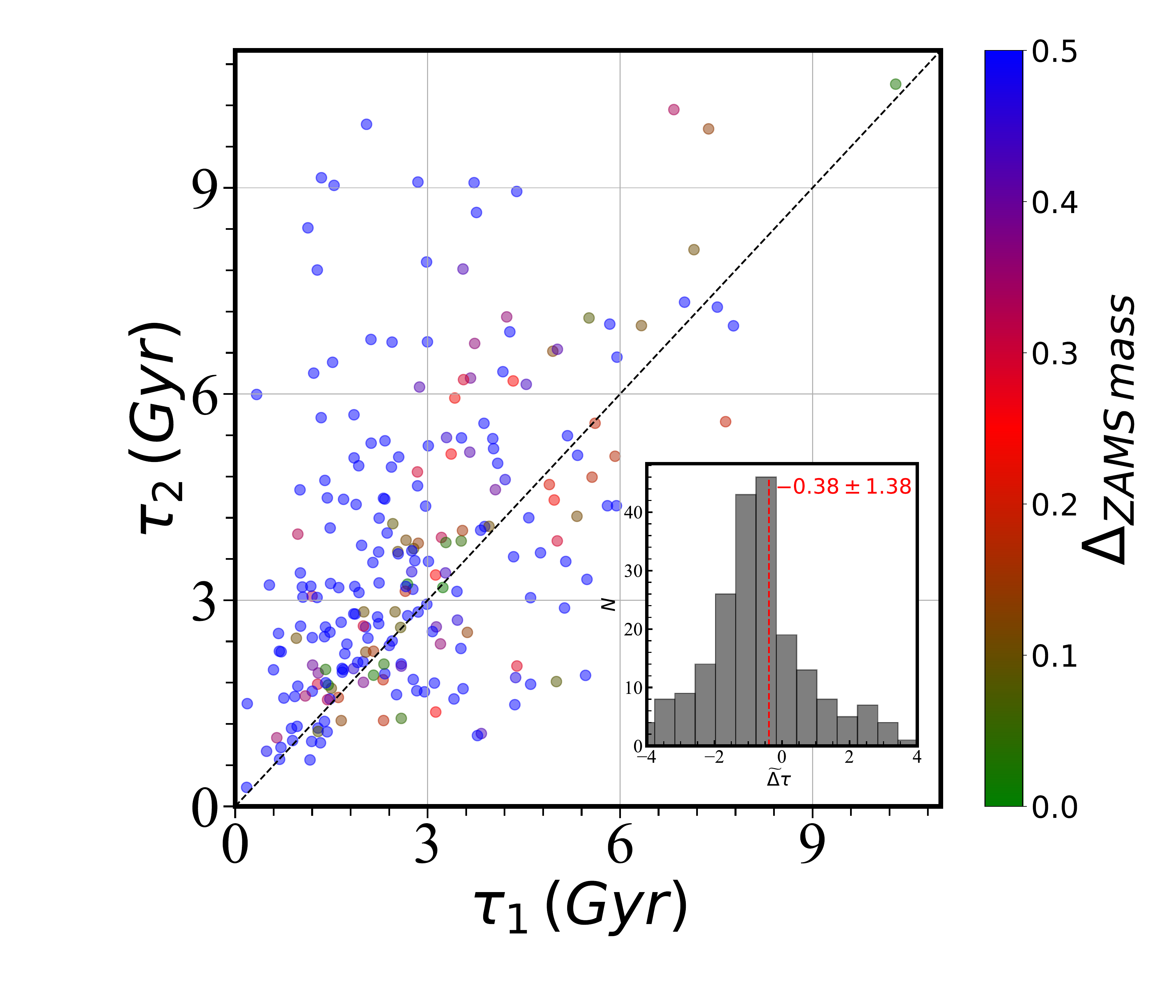}
\caption{Age comparison of the two components of 236 WD-WD binaries, colour-coded by mass differences $\rm \Delta_{mass}$ ($\equiv \rm mass_1-mass_2$), where $\rm mass_1$ and $\rm mass_2$ represent the mass of the primary and secondary components, respectively.  The insert sub-panel displays the histogram of the error-weighted difference in age between the two components in WD-WD candidate binaries, e,g. $\widetilde\Delta\tau$=($\tau_1$-$\tau_2$)/$\sqrt{\sigma^2_{\tau_1}+\sigma^2_{\tau_2}}$, where $\tau_1$, $\tau_2$, $\sigma_{\tau_1}$, and $\sigma_{\tau_2}$ represents the age and age uncertainties of the primary and secondary components, respectively. The red dashed line marks the mean value of $\widetilde\Delta\tau$.}
\label{fig:Age-difference}
\end{figure}

In the second approach, we used a subsample of WD-WD candidates for the age cross-validation, those with no significant contamination probability, and with good age constraints: We eliminate those with $\log(s/AU) > 4.6$, or $\sigma_{\Delta\mu}>1.0$ ($\sigma_{\Delta\mu}$ is defined in Section 2.1 of T20) leaving us with 236  WD-WD binaries. Using BASE-9, we determine the ages for these cWDs. Figure \ref{fig:Age-difference} shows the age of the two components for 236 WD-WD binaries, colour-coded by mass differences $\rm \Delta_{mass}$ ($ \equiv \rm mass_{1} - mass_{2}$), where $\rm mass_{1}$ and $\rm mass_{2}$ represent the mass of the primary and secondary components, respectively. The insert sub-plot represents the  histograms of the error-weighted age differences ($\widetilde\Delta\tau$=($\tau_1$-$\tau_2$)/$\sqrt{\sigma^2_{\tau_1}+\sigma^2_{\tau_2}}$) between the primary and secondary components of WD-WD candidate binaries, where $\tau_1$, $\tau_2$, $\sigma_{\tau_1}$, and $\sigma_{\tau_2}$ represents the ages and their uncertainties of the two components, respectively. The mean (median) value -0.38 (-0.72) $\pm$ 1.38, marked by the red dashed line in Figure \ref{fig:Age-difference}) of $\widetilde\Delta\tau$ suggests that the age determination has a systematic dependence on the ZAMS mass, this is presumably because the IFMR is systematically wrong. The ages of the two components are consistent within 1-2 $\sigma$. Considering that the photometric data is independently observed for each member cWD, such a small age difference indicates that the stellar ages obtained with BASE-9 are reliable. Note that this set of WD-WD binaries has vastly more potential to constrain the IFMR.

\begin{table*} 
\centering
\tabletypesize{\scriptsize}
 \begin{threeparttable}
\caption{Catalog description of the field MS components from the primary dataset} \label{tab:s<4.4}
\begin{tabular}{l|l|l}
\hline\hline
Column & units & Description \\ \hline
\texttt{source\_id}                          &         & {\it Gaia} source id (int64); MS  \\
\texttt{dec}                                 & deg     & declination from Gaia; MS  \\
\texttt{ra}                                  & deg     & right ascension from Gaia; MS  \\
\texttt{pairdistance}                        & deg     & angular separation between MS and WD  \\
\texttt{parallax}                            & mas     & parallax; MS  \\
\texttt{parallax\_over\_error}               &         & parallax divided by its error; MS  \\
\texttt{phot\_g\_mean\_mag}                  & mag     & G-band mean magnitude (Vega scale); MS  \\
\texttt{phot\_g\_mag\_error}                  & mag     & G-band magnitude uncertainty; MS  \\
\texttt{phot\_bp\_mean\_mag}                 & mag     & integrated BP mean magnitude; MS  \\
\texttt{phot\_bp\_mag\_error}                  & mag     & BP-band magnitude uncertainty; MS  \\
\texttt{phot\_rp\_mean\_mag}                 & mag     & integrated RP mean magnitude; MS  \\
\texttt{phot\_rp\_mag\_error}                  & mag     & RP-band magnitude uncertainty; MS  \\
\texttt{s\_AU}                               & AU      & projected physical separation between two components \\
\texttt{pmdec}                               & mas\,yr$^{-1}$ & proper motion in the declination direction; MS  \\
\texttt{pmdec\_error}                        & mas\,yr$^{-1}$ & standard error of proper motion in the declination direction; MS \\
\texttt{pmra}                                & mas\,yr$^{-1}$ & proper motion in right ascension direction; i.e., $\mu_{\alpha}^{*} = \mu_{\alpha}\cos{\delta}$; MS  \\
\texttt{pmra\_error}                         & mas\,yr$^{-1}$ & standard error of proper motion in right ascension direction; MS \\
\texttt{radial\_velocity}                    & km\,s$^{-1}$ & spectroscopic barycentric radial velocity; MS \\
\texttt{radial\_velocity\_error}             & km\,s$^{-1}$ & standard error of spectroscopic barycentric radial velocity; MS  \\
\texttt{mn\_age}               & Gyr     & posterior median WD age\\
\texttt{mn\_preAge}               & Gyr     & posterior mean WD precusor's age\\ 
\texttt{mn\_coolAge}               & Gyr     & posterior mean WD cooling age\\
\texttt{mn\_[Fe/H]}               & dex     & posterior mean [Fe/H]\\
\texttt{mn\_Distmod}               & mag     & posterior mean distance modulus\\
\texttt{mn\_$A_V$}               & mag     & posterior mean WD extinction in V-band\\
\texttt{st\_age}               & Gyr     & posterior standard deviation WD age\\
\texttt{st\_preAge}               & Gyr      & posterior standard deviation WD precusor's age\\
\texttt{st\_coolAge}               & Gyr    & posterior standard deviation WD cooling age\\
\texttt{st\_[Fe/H]}               & dex     & posterior standard deviation WD [Fe/H]\\
\texttt{st\_Distmod}               & mag     & posterior standard deviation distance modulus\\
\texttt{st\_$A_V$}               & mag    & posterior standard deviation WD extinction in V-band\\
\texttt{Nband} &  & number of photometry bands used in BASE-9 fitting\\
\texttt{P\_contamination} & & contamination probability by random alignments\\
\hline
\hline
\end{tabular}
 \end{threeparttable}
\end{table*}

\begin{table}
\centering
\caption{}
\label{tab:Age}
\begin{tabular}{ l|l|l|l|l|l} 
\hline
\hline
\multicolumn{2}{c|}{WDs}&Clusters&\multicolumn{2}{c|}{$\log(\tau/yr)$}\\
\hline
 RA& DEC & Name & Ours  &Reference&Nband\\
\hline
129.939& 20.004 & Praesepe& $8.89_{-0.07}^{+0.10}$ &$ 8.85_{-0.06}^{+0.08}$&7\\
127.166 & 19.728 & Praesepe& $8.81_{-0.03}^{+0.06}$ &$ 8.85_{-0.06}^{+0.08}$&7\\
130.055 & 18.723 & Praesepe& $8.82_{-0.03}^{+0.05}$ &$ 8.85_{-0.06}^{+0.08}$&7\\
130.841 & 20.725 & Praesepe& $8.85_{-0.03}^{+0.04}$ &$ 8.85_{-0.06}^{+0.08}$&7\\
129.946 & 19.769 & Praesepe& $8.88_{-0.03}^{+0.03}$ &$ 8.85_{-0.06}^{+0.08}$&7\\
131.507 & 18.513 & Praesepe& $8.85_{-0.03}^{+0.05}$ &$ 8.85_{-0.06}^{+0.08}$&7\\
289.057 & -16.337 & NGC 6774& $9.36_{-0.29}^{+0.33}$ &$ 9.30_{-0.06}^{+0.08}$&7\\
288.765 & -16.061 & NGC 6774& $9.32_{-0.24}^{+0.38}$ &$ 9.30_{-0.06}^{+0.08}$&7\\
289.684 & -15.898  & NGC 6774& $9.24_{-0.19}^{+0.30}$ &$ 9.30_{-0.06}^{+0.08}$&7\\
184.734 & 25.765  & ComaBer& $8.76_{-0.02}^{+0.02}$ &$ 8.81_{-0.06}^{+0.08}$&7\\
132.832 & 11.811  & NGC 2682& $9.47_{-0.03}^{+0.08}$ &$ 9.54_{-0.06}^{+0.08}$&7\\
\hline
\hline
\end{tabular}
\end{table}

\section{Conclusion}
\label{sect:conclution}

In this work, we select 9589 MS-WD and 307 WD-WD candidate binaries from the catalog released by T20, which is a catalog of 807,611 binaries by searching from {\it Gaia} DR2 in a range of distance $<$ 4.0\,kpc and the projected separation $s<$ 1.0\,pc. Due to lack of the line-of-sight velocities for most of the candidates, the sample built in this work inherits the high contamination rate from its parent catalog at the large separations. Particularly, at $d>0.5$\,kpc, greater than 80\% of candidate binary are dominated by the chance alignments. According to the contamination rates, we selected 4050 MS-WD binaries with contamination rates $<20\%$.

We use an open source software suite, BASE-9 \citep{Hipple_2014}, to constrain the ages, ZAMS mass, $A_V$ and distance modulus for the 4050 MS-WD candidate binaries. Discarding the cases (499 cWDs) with very poor age convergences, we finally obtain 903 ages with relative uncertainties $\sigma_{\tau}/\tau < 10$\%, 665 ages with 10\% $<\sigma_{\tau}/\tau < $20\%, 655 ages with 20\% $<\sigma_{\tau}/\tau < $ 30\%, and 894 ages with 30\% $<\sigma_{\tau}/\tau < $ 50\%. The remaining 434 cWDs have $\sigma_{\tau}/\tau > $50\%. The large age uncertainty is caused by the low precision parallax and/or photometric measurements for the distant cWDs. The vast majority of ages are in the range of 1 to 8\,Gyr. 

We adopt two independent methods to validate the ages derived by BASE-9. One is by means of a few well known clusters, the other is to use the WD-WD catalog for cross-validation. Both the methods demonstrate that our ages are well derived by BASE-9 in both the accuracy and precision. The age uncertainties can be improved by a factor of two or three relative to other typical approaches of age estimation. 

Under a safe assumption that the components are co-eval in a binary system, we can obtain the age of a MS star from its companion WD. However, our MS-WD sample is contaminated by the chance alignments, particularly at large separations. Statistically speaking, the age can be safely equivalent with its companion WD for the MS-WD candidate binaries with contaminated by a low rate, e.g., $<20\%$. We build a catalog that includes the ages and other parameters (see Table \ref{tab:s<4.4}) of the 3551 field MS components. We also provide a catalog which includes the ages and other parameters of the 3551 MS-WD and 236 WD-WD binaries (see the Appendix). These two catalogs will be released on-line. 

We expect to enlarge the MS-WD binary catalog to a great extent by combining the Gaia DR2 (or even DR3) and the new faint proper motions catalog GPS1+ \citep{Tian_2020_GPS1+} in the future. Gaia DR2 provides more than 1.3 billion stars brighter than 20.7\,mag in the G-band which have measured positions, proper motions, parallaxes and colors with unprecedented precision. GPS1+ released good proper motions for almost 400 million stars fainter than 20th\,mag in the r-band. Like \citet{Fouesneau2019} has done, we intend to select more faint cWDs from GPS1+, and obtain the parallaxes of the companion MS from Gaia DR2 and photometry from PS1. 

\acknowledgements
We thank Jie Su, Jingkun Zhao, and Xianfei Zhang for helpful discussions, and acknowledges the National Natural Science Foundation of China (NSFC) under grants 11873034, U1731108, and U1731124. This material is partly based upon work supported by the National Science Foundation under Grant No.\ AST-1715718, and by the European Research Council under the European Unions Seventh Framework Programme (FP 7) ERC Grant Agreement n. [321035]. This work has made use of data from the European Space Agency (ESA) mission
{\it Gaia} (\url{https://www.cosmos.esa.int/gaia}), processed by the {\it Gaia}
Data Processing and Analysis Consortium (DPAC, \url{https://www.cosmos.esa.int/web/gaia/dpac/consortium}). Funding for the DPAC has been provided by national institutions, in particular the institutions
participating in the {\it Gaia} Multilateral Agreement.

This research made use of TOPCAT, an interactive graphical viewer and editor for tabular data \citep{Taylor2005}, matplotlib, a Python library for publication quality graphics \citep{Hunter:2007}, NumPy \citep{van2011}.
 

\appendix

We finally build a total catalog including both the 3551 MS-WD pairs with contaminated rate $<20\%$ and 236 WD-WD pairs datasets, and includes the ages and other parameters for all the 3787 pairs as described in Table \ref{tab:all}.

 \begin{longtable*} {p{3.5cm} | p{1.2cm}| p{11.5cm}}
\caption{Catalog description} \\ \hline\hline
\label{tab:all}
Column & units & Description \\ \hline
\texttt{source\_id}                          &         & {\it Gaia} source id (int64); star 1  \\
\texttt{source\_id2}                         &         & {\it Gaia} source id (int64); star 2  \\
\texttt{ra}                                  & deg     & right ascension from Gaia; star 1  \\
\texttt{ra2}                               & deg     & right ascension from Gaia; star 2  \\
\texttt{dec}                                 & deg     & declination from Gaia; star 1  \\
\texttt{dec2}                                & deg     & declination from Gaia; star 2  \\
\texttt{parallax}                            & mas     & parallax; star 1  \\
\texttt{parallax2}                           & mas     & parallax; star 2  \\
\texttt{parallax\_over\_error}               &         & parallax divided by its error; star 1  \\
\texttt{parallax\_over\_error2}              &         & parallax divided by its error; star 2 \\
\texttt{phot\_g\_mean\_mag}                  & mag     & G-band mean magnitude (Vega scale); star 1  \\
\texttt{phot\_g\_mean\_mag2}                 & mag     & G-band mean magnitude (Vega scale); star 2  \\
\texttt{phot\_g\_mag\_error}                  & mag     & G-band magnitude uncertainty; star 1  \\
\texttt{phot\_g\_mag\_error2}                  & mag     & G-band magnitude uncertainty; star 2  \\
\texttt{phot\_bp\_mean\_mag}                 & mag     & integrated BP mean magnitude; star 1  \\
\texttt{phot\_bp\_mean\_mag2}                & mag     & integrated BP mean magnitude; star 2  \\
\texttt{phot\_bp\_mag\_error}                  & mag     & BP-band magnitude uncertainty; star 1  \\
\texttt{phot\_bp\_mag\_error2}                  & mag     & BP-band magnitude uncertainty; star 2  \\
\texttt{phot\_rp\_mean\_mag}                 & mag     & integrated RP mean magnitude; star 1  \\
\texttt{phot\_rp\_mean\_mag2}                & mag     & integrated RP mean magnitude; star 2  \\
\texttt{phot\_rp\_mag\_error}                  & mag     & RP-band magnitude uncertainty; star 1  \\
\texttt{phot\_rp\_mag\_error2}                  & mag     & RP-band magnitude uncertainty; star 2  \\
\texttt{pmra}                                & mas\,yr$^{-1}$ & proper motion in right ascension direction; i.e., $\mu_{\alpha}^{*} = \mu_{\alpha}\cos{\delta}$; star 1  \\
\texttt{pmra2}                               & mas\,yr$^{-1}$ & proper motion in right ascension direction; i.e., $\mu_{\alpha}^{*} = \mu_{\alpha}\cos{\delta}$; star 2  \\
\texttt{pmra\_error}                         & mas\,yr$^{-1}$ & standard error of proper motion in right ascension direction; star 1  \\
\texttt{pmra\_error2}                        & mas\,yr$^{-1}$ & standard error of proper motion in right ascension direction; star 2  \\
\texttt{pmdec}                               & mas\,yr$^{-1}$ & proper motion in the declination direction; star 1  \\
\texttt{pmdec2}                              & mas\,yr$^{-1}$ & proper motion in the declination direction; star 2  \\
\texttt{pmdec\_error}                        & mas\,yr$^{-1}$ & standard error of proper motion in the declination direction; star 1  \\
\texttt{pmdec\_error2}                       & mas\,yr$^{-1}$ & standard error of proper motion in the declination direction; star 2  \\
\texttt{radial\_velocity}                    & km\,s$^{-1}$ & spectroscopic barycentric radial velocity; star 1  \\
\texttt{radial\_velocity2}                   & km\,s$^{-1}$ & spectroscopic barycentric radial velocity; star 1  \\
\texttt{radial\_velocity\_error}             & km\,s$^{-1}$ & standard error of spectroscopic barycentric radial velocity; star 1  \\
\texttt{radial\_velocity\_error2}            & km\,s$^{-1}$ & standard error of spectroscopic barycentric radial velocity; star 2  \\

\texttt{gmag}                    & mag            &  g magnitude; star1 \\
\texttt{gmag2}                    & mag            &  g magnitude; star2 \\
\texttt{rmag}                    & mag            &  r magnitude; star1 \\
\texttt{rmag2}                    & mag            &  r magnitude; star2 \\
\texttt{imag}                    & mag            &  i magnitude; star1 \\
\texttt{imag2}                    & mag            &  i magnitude; star2 \\
\texttt{zmag}                    & mag            &  z magnitude; star1 \\
\hline
\newpage
\hline
\hline
Column & units & Description \\
 \hline
 \texttt{zmag2}                    & mag            &  z magnitude; star2 \\
\texttt{e\_gmag}            & mag          &  g uncertainty; star1 \\
\texttt{e\_gmag2}            & mag          &  g uncertainty; star2 \\
\texttt{e\_rmag}            & mag          &  r uncertainty; star1 \\
\texttt{e\_rmag2}            & mag          &  r uncertainty; star2 \\
\texttt{e\_imag}            & mag          &  i uncertainty; star1 \\
\texttt{e\_imag2}            & mag          &  i uncertainty; star2 \\
\texttt{e\_zmag}            & mag          &  z uncertainty; star1 \\
\texttt{e\_zmag2}            & mag          &  z uncertainty; star2 \\

\texttt{mn\_age}               & Gyr     & posterior median WD age; star1\\
\texttt{mn\_age2}               & Gyr     & posterior median WD age; star2\\
\texttt{mn\_ZAMS\_mass}               & \Msun\     & posterior mean WD ZAMS mass; star1 \\
\texttt{mn\_ZAMS\_mass2}               & \Msun\     & posterior mean WD ZAMS mass; star2 \\

\texttt{mn\_preAge}               & Gyr     & posterior mean WD precusor's age; star1\\ 
\texttt{mn\_preAge2}               & Gyr     & posterior mean WD precusor's age; star2\\ 
\texttt{mn\_coolAge}               & Gyr     & posterior mean WD cooling age; star1\\
\texttt{mn\_coolAge2}               & Gyr     & posterior mean WD cooling age; star2\\
\texttt{mn\_[Fe/H]}               & dex     & posterior mean [Fe/H]; star1\\
\texttt{mn\_[Fe/H]2}               & dex     & posterior mean [Fe/H]; star2\\
\texttt{mn\_Distmod}               & mag     & posterior mean distance modulus; star1 \\
\texttt{mn\_Distmod2}               & mag     & posterior mean distance modulus; star2 \\
\texttt{mn\_$A_V$}               & mag     & posterior mean WD extinction in V-band; star1\\
\texttt{mn\_$A_V2$}               & mag     & posterior mean WD extinction in V-band; star2\\
\texttt{st\_age}               & Gyr     & posterior standard deviation WD age; star1\\
\texttt{st\_age2}               & Gyr     & posterior standard deviation WD age; star2\\
\texttt{st\_ZAMS\_mass}               & \Msun\      & posterior standard deviation WD ZAMS mass; star1\\
\texttt{st\_ZAMS\_mass2}               & \Msun\      & posterior standard deviation WD ZAMS mass; star2\\
\texttt{st\_preAge}               & Gyr      & posterior standard deviation WD precusor's age; star1\\
\texttt{st\_preAge2}               & Gyr      & posterior standard deviation WD precusor's age; star2\\
\texttt{st\_coolAge}               & Gyr    & posterior standard deviation WD cooling age; star1\\
\texttt{st\_coolAge2}               & Gyr    & posterior standard deviation WD cooling age; star2\\
\texttt{st\_[Fe/H]}               & dex     & posterior standard deviation WD [Fe/H]; star1\\
\texttt{st\_[Fe/H]2}               & dex     & posterior standard deviation WD [Fe/H]; star2\\
\texttt{st\_Distmod}               & mag     & posterior standard deviation distance modulus; star1\\
\texttt{st\_Distmod2}               & mag     & posterior standard deviation distance modulus; star2\\
\texttt{st\_$A_V$}               & mag    & posterior standard deviation WD extinction in V-band; star1\\
\texttt{st\_$A_V2$}               & mag    & posterior standard deviation WD extinction in V-band; star2\\
\texttt{Nband} &  & type of photometric band which BASE-9 fitted, 7 bands and 3 bands are assigned with 7 and 3, respectively; star1\\
\texttt{Nband2} &  & type of photometric band which BASE-9 fitted, 7 bands and 3 bands are assigned with 7 and 3, respectively; star2\\
\texttt{s\_AU}                               & AU      & projected physical separation between two stars \\
\texttt{pairdistance}                        & deg     & angular separation between star 1 and star 2  \\
\texttt{binary\_type}   &       & type of wide binary. Different types are assigned different integer identifiers: MS-WD and WD-WD are assigned with 1 and 2, respectively. \\
\texttt{P\_contamination} & & contamination probability by the random alignments\\
\hline
\hline

 \end{longtable*}

\begin{flushleft}
Note: Each row in the catalog corresponds to a single binary; "star 1" and "star 2" designation in each binary are arbitrary.
\end{flushleft}
\label{sect:all}

\end{CJK}
\end{document}